\documentclass[10pt,conference]{IEEEtran}
\usepackage[utf8]{inputenc}
\IEEEoverridecommandlockouts

\usepackage{cite}
\usepackage{amsmath,amssymb,amsfonts}
\usepackage{graphicx}
\usepackage{textcomp}
\usepackage{xcolor}
\usepackage{xspace}
\usepackage{listings}
\usepackage{amssymb}
\usepackage{color}
\usepackage{multirow}
\usepackage{dblfloatfix}
\usepackage{float}
\usepackage{lscape}
\usepackage{xparse}
\usepackage{placeins}
\usepackage{caption}
\usepackage{subcaption}
\usepackage{enumitem}
\usepackage{array, makecell}
\usepackage{wrapfig}
\usepackage{colortbl}
\usepackage{tabularx}
\usepackage{booktabs}
\usepackage{courier}
\usepackage{ragged2e}
\usepackage{amsthm}
\usepackage{hyperref}
\usepackage{comment}
\usepackage{pdflscape}
\usepackage{afterpage}
\usepackage{fontawesome}
\usepackage{pifont}
\usepackage{lscape}

\usepackage{enumitem}
\usepackage{balance}

\usepackage[linesnumbered,ruled,vlined]{algorithm2e}
\usepackage{amsmath}
\usepackage{amssymb}
\usepackage{listings}
\usepackage{inconsolata}

\setlist[itemize]{noitemsep, topsep=4pt}
\definecolor{lightgray}{HTML}{f6f6f6}
\definecolor{darkgray}{rgb}{.4,.4,.4}
\definecolor{darkblue}{HTML}{1b4db3}
\definecolor{brickred}{HTML}{b04f4f}
\definecolor{purple}{rgb}{0.65, 0.12, 0.82}
\definecolor{diffadd}{HTML}{288f26}
\definecolor{diffrmbg}{HTML}{ffebe9}
\definecolor{diffaddbg}{HTML}{e6ffeb}
\definecolor{diffremove}{HTML}{de4f54}
\definecolor{carrotorange}{rgb}{0.8, 0.33, 0.0}
\definecolor{highlight}{HTML}{fefbc2}
\lstdefinelanguage{JavaScript}{
  keywords={typeof, new, true, false, catch, function, return, null, catch, switch, var, const, let, extends, if, in, while, do, else, case, break, async, await,of},
  keywordstyle=\color{darkblue}\bfseries,
  ndkeywords={class, export, boolean, throw, implements, import, this, setTimeout},
  ndkeywordstyle=\color{brickred}\bfseries,
  identifierstyle=\color{black},
  sensitive=false,
  comment=[l]{//},
  morecomment=[f][\color{diffadd}\bfseries]{+\ },
  morecomment=[s]{/*}{*/},
  morecomment=[f][\color{diffremove}\bfseries]{- },
  commentstyle=\color{violet}\ttfamily,
  stringstyle=\color{carrotorange}\ttfamily,
  morestring=[b]',
  morestring=[b]"
}
\lstdefinelanguage{Java}{
	keywords={new, true, false, catch, void, public, return, null, catch, int},
	keywordstyle=\color{darkblue}\bfseries,
	ndkeywords={class, export, boolean, throw, implements, import, this, setTimeout},
	ndkeywordstyle=\color{brickred}\bfseries,
	identifierstyle=\color{black},
	sensitive=false,
	comment=[l]{//},
	morecomment=[f][\color{diffadd}\bfseries]{+\ },
	morecomment=[s]{/*}{*/},
	morecomment=[f][\color{diffremove}\bfseries]{- },
	commentstyle=\color{violet}\ttfamily,
	stringstyle=\color{carrotorange}\ttfamily,
	morestring=[b]',
	morestring=[b]"
}

\theoremstyle{definition}
\newtheorem{definition}{Definition}
\newcommand{\header}[1]{\par\smallskip\noindent\textbf{#1.}}

\def\BibTeX{{\rm B\kern-.05em{\sc i\kern-.025em b}\kern-.08em
    T\kern-.1667em\lower.7ex\hbox{E}\kern-.125emX}}
    
\newboolean{showcomments}
\setboolean{showcomments}{true}
\ifthenelse{\boolean{showcomments}}
{
	\definecolor{myyellow}{RGB}{255, 228, 26}
	\definecolor{myblue}{RGB}{50, 50, 220}
	\newcommand{\nb}[2]{
		{\sf
			\fcolorbox{myyellow}{yellow}{\scriptsize\textbf{#1}}%
			$\blacktriangleright$%
			{\color{myblue}\fontsize{7pt}{8pt}\selectfont\textbf{#2}}%
		}%
	}
}
{
	\newcommand{\nb}[2]{}
}

\newcommand{\ali}[1]{\nb{Ali}{#1}}

\newcommand{\benchmark}{\textsc{Birch}\xspace}

\newcommand{\benchmarkdataset}{\textsc{Hunk4J}\xspace}
\newcommand{\hunkdivergence}{\emph{hunk divergence}\xspace}
\newcommand{\proximityclass}{\emph{spatial proximity}\xspace}

\newcommand{\code}[1]{{\small\ttfamily\texttt{#1}}}

\begin{document}

\title{
% Multi-Hunk Divergence: Characterization and Repair with LLMs
% Multi-Hunk Divergence, Proximity, and LLM Repair Challenges
% Multi-Hunk Patches: Divergence, Proximity and the Limits of LLM
Characterizing Multi-Hunk Patches: Divergence, Proximity, and LLM Repair Challenges % [ASE Title]
%Hunk Divergence and Proximity in LLM-based Multi-Hunk Patch Generation
%HunkBench: Understanding Edit Divergence in Multi-Hunk Repair}
%HunkBench: Understanding Divergence for Code Repair}
%Understanding Hunk Divergence for Code Repair}
%Understanding Hunk Divergence for Multi-location Patch}
%Understanding Hunk Divergence for Multi-location Repair}
%Repairing Multi-Hunk Bugs with LLMs: Benchmark, Characterization, and Divergence Metric}
%Multi-hunk Patch Divergence: Characterization and Repair}
%Understanding Hunk Divergence in Multi-location Repair with LLMs}
%Code Agents and Hunk Divergence: Toward Effective Multi-location Repair}
%Code Agents, LLMs, and Hunk Divergence in Multi-location Repair}
%HunkBench: Code Agents, LLMs, and Hunk Divergence in Multi-location Repair}
%HunkBench: Evaluating LLMs and Code Agents on Multi-Hunk Repair through Hunk Divergence Analysis
}

% \author{\IEEEauthorblockN{Anonymous Authors}}
\makeatletter
\newcommand{\linebreakand}{%
  \end{@IEEEauthorhalign}%
  \hfill\mbox{}\par%
  \mbox{}\hfill\begin{@IEEEauthorhalign}%
}
\makeatother

\author{%
  \IEEEauthorblockN{Noor Nashid}
  \IEEEauthorblockA{%
    \textit{University of British Columbia}\\
    Vancouver, Canada\\
    nashid@ece.ubc.ca%
  }
  \and
  \IEEEauthorblockN{Daniel Ding}
  \IEEEauthorblockA{%
    \textit{University of British Columbia}\\
    Vancouver, Canada\\
    dyxd2003@ece.ubc.ca%
  }
  \and
  \IEEEauthorblockN{Keheliya Gallaba}
  \IEEEauthorblockA{%
    \textit{Queen’s University}\\
    Kingston, Canada\\
    gallabak@sigsoft.org%
  }
  \linebreakand
  \IEEEauthorblockN{Ahmed E. Hassan}
  \IEEEauthorblockA{%
    \textit{Queen’s University}\\
    Kingston, Canada\\
    ahmed@cs.queensu.ca%
  }
  \and
  \IEEEauthorblockN{Ali Mesbah}
  \IEEEauthorblockA{%
    \textit{University of British Columbia}\\
    Vancouver, Canada\\
    amesbah@ece.ubc.ca%
  }
}
\maketitle

\thispagestyle{plain}
\pagestyle{plain}

\begin{abstract}
%\ali{abstract needs a re-write}

% \keheliya{spatially}
Multi-hunk bugs, where fixes span disjoint regions of code, are common in practice, yet remain underrepresented in automated repair. Existing techniques and benchmarks predominantly target single-hunk scenarios, overlooking the added complexity of coordinating semantically related changes across the codebase. In this work, we characterize \benchmarkdataset, a dataset of multi-hunk patches derived from 372 real-world defects. We propose \hunkdivergence, a metric that quantifies the variation among edits in a patch by capturing lexical, structural, and file-level differences, while incorporating the number of hunks involved. We further define \proximityclass, a classification that models how hunks are spatially distributed across the program hierarchy. Our empirical study spanning six LLMs reveals that model success rates decline with increased divergence and spatial dispersion. Notably, when using the LLM alone, no model succeeds in the most dispersed \emph{Fragment} class. 
These findings highlight a critical gap in LLM capabilities and motivate \emph{divergence-aware} repair strategies.
%These insights expose limitations in current approaches and motivate the development of \emph{divergence-aware} repair techniques.
\end{abstract}


% !TEX root =  ../main.tex
\section{Introduction}

Automated Program Repair (APR) has received significant attention from the research community~\cite{yasunaga:drrepair:icml20, cure-program-repair-icse-2021, codebert-java-sstubs-21, recoder-fse-2021, rewardrepair-icse-2022, hata2019learning, watson:lit-review-dl4se:tosem:22, le:srevey-dl4se:csur20, tufano:tosem:19, reptory, coconut, sequencer, zhang:apr-survey:tosem23, huang:apr-survey:acm24, deepdelta, hoppity, glance, katana}. APR techniques have progressed from heuristic-based methods~\cite{legoues:apr:acm19, legoues:genprog:tse11, weimer:everaging-program-equivalence:ase13, kim:par-template-based-apr:icse13} to deep learning approaches that learn repair patterns from large codebases~\cite{deepdelta, hoppity, glance, reptory, katana}, achieving promising results on single-hunk bugs~\cite{hoppity, glance, katana, reptory, rewardrepair-icse-2022, recoder-fse-2021, li:dlfix:icse20, xuan:nopol:tse16, jiang:simfix:issta18, zhu:tare:icse23, yang:transplantfix:ase22}. Despite recent progress, most APR research remains focused on single-hunk scenarios~\cite{hoppity, xuan:nopol:tse16, glance, katana, reptory}, overlooking \emph{multi-hunk bugs}, defects that require edits across disjoint code regions. These are both common in practice~\cite{hercules} and harder to fix due to their semantic interdependencies. While the problem has been acknowledged~\cite{hercules}, only a few studies have directly addressed it~\cite{mechtaev:angelix-multi-hunk:icse16, hercules, yuan:arjae:19, wong:varfix:fse21, dear, iter}, and they have demonstrated limited effectiveness.

Recent advancements in large language models (LLMs) have shown promise in transforming several software engineering tasks, including program repair~\cite{cedar}. Trained on massive datasets comprising code and natural language, LLMs such as GPT-4 and LLaMA have demonstrated the ability to understand and generate code~\cite{xia:alpharepair:fse22, hossain:llm-bug-localization-and-repair:fse24, zhang:critical-review-chatgpt-and-apr:arxiv23, cedar}. While these models have already been employed to address single-hunk bugs with notable success, their potential to handle multi-hunk code repair remains underexplored.

% Early studies have examined LLM-based multi-hunk repair using tools such as AlphaRepair~\cite{xia:alpharepair:fse22}, RepairAgent~\cite{repair-agent}, and fine-tuning strategies~\cite{huang:empirical-fine-tuning-for-apr:ase23, yang:fine-tuning-for-repair-with-llm:arxiv24}. 
% Although these efforts demonstrate that LLMs can generalize beyond single-hunk scenarios, they also reveal \ali{did they explicitly say this? or are we anticipating these limitations?} limitations in handling semantically interdependent edits across dispersed code locations~\cite{xin:empirical-study-hunk:fse24}. 
% \ali{did they explicitly say this? or are we anticipating these limitations?}
% Since they do not explicitly mention LLMs, we have rewritten the sentence to reflect this distinction accurately.
% Although these efforts demonstrate that LLMs can generalize beyond single-hunk scenarios, findings from~\cite{xin:empirical-study-hunk:fse24} highlight challenges in handling semantically interdependent edits across dispersed code locations, challenges that likely extend to LLM-based techniques as well.
%The absence of standardized benchmarks and unified evaluation criteria for multi-hunk repair hinders the interpretability and comparability of these findings.
Early studies on LLM-based multi-hunk repair include tools such as AlphaRepair~\cite{xia:alpharepair:fse22}, RepairAgent~\cite{repair-agent}, and fine-tuning strategies~\cite{huang:empirical-fine-tuning-for-apr:ase23, yang:fine-tuning-for-repair-with-llm:arxiv24}. Although these approaches demonstrate that LLMs can extend beyond single-hunk scenarios, the challenges of reasoning about semantically interdependent edits distributed across the codebase remain unaddressed.
% As a result, a systematic characterization of the factors influencing LLM effectiveness on multi-hunk repair remains an open problem.

We identify four challenges limiting progress. First, \emph{dataset variation} impedes consistency: prior studies evaluate on synthetic benchmarks such as LMDefects~\cite{fan:code-repair-with-llm:icse23}, algorithmically curated tasks (EvalRepair-C++/Java)~\cite{yang:fine-tuning-for-repair-with-llm:arxiv24}, and real-world datasets such as Defects4J~\cite{srepair:arxiv24}, often under varying assumptions, input formats, and test harnesses. Second, \emph{bug-type ambiguity} obscures evaluation: multi-hunk bugs span a wide spectrum, from edits in non-contiguous regions within a single function to interdependent changes across multiple files. Yet prior work does not systematically characterize this \emph{heterogeneity}, often conflating distinct forms of multi-location repair such as multi-line, multi-function, and multi-file edits. Third, \emph{reporting inconsistencies} hinder meaningful attribution: existing techniques often report aggregate metrics that combine single- and multi-hunk results~\cite{yang:fine-tuning-for-repair-with-llm:arxiv24, repair-agent}, making it difficult to isolate and evaluate their effectiveness specifically on multi-hunk repairs. Fourth, existing evaluations lack \emph{patch complexity-aware metrics} that quantify differences in hunks within a patch. %This gap highlights the need for a systematic approach to multi-hunk repair that explicitly models these forms of \emph{divergence}, factors central to understanding the complexity of multi-hunk patches.

%To address these gaps, we propose a framework for the systematic analysis of multi-hunk bugs. First, we introduce \emph{hunk divergence}, a metric that quantifies the internal variation within a patch by measuring lexical, structural, and file-level distance among hunks, while accounting for the number of hunks and the dispersion across files. Second, we define \emph{spatial proximity}, a classification that reflects hunk dispersion across methods, files, and packages in the codebase. Together, these abstractions allow us to analyze patch complexity systematically.

To address these challenges, we (1) propose \emph{hunk divergence}, a metric that quantifies the internal variation within a multi-hunk patch by measuring lexical, structural, and file-level distance among hunks, while accounting for the number of hunks and the dispersion across files; (2) define \emph{spatial proximity} as a classification scheme that assigns each patch to one of five categories---Nucleus, Cluster, Orbit, Sprawl, or Fragment---reflecting increasing levels of dispersion across the program hierarchy, including methods, files, and packages; (3) characterize multi-hunk patches using these abstractions to analyze hunk divergence and inter-hunk relationships; (4) develop \benchmark, a platform for standardized evaluation of LLM-based multi-hunk repair that supports reproducible comparisons across models and configurations; and (5) conduct a large-scale empirical study, examining the impact of retrieval strategies, context granularity, and feedback on repair success and cost under varying degrees of divergence and dispersion.

In this work, we make the following contributions:

\begin{itemize}    
    \item We propose \emph{hunk divergence}, a metric that quantifies intra-patch variation across lexical, structural, and file-level distance, and \emph{spatial proximity}, a classification of multi-hunk patches based on their spatial layout in the codebase.

    \item We introduce \benchmarkdataset, a benchmark of 372 real-world multi-hunk bugs. We also present the first systematic characterization of multi-hunk bugs, focusing on their hunk diversity and proximity.  

    \item We present \benchmark, a platform for evaluating the capabilities of LLMs on multi-hunk bugs. Our empirical study spans 6 LLMs, various modes including 2 code representations, and 3 retrieval strategies, while also examining the role of contextual scope and incorporating feedback from compilation errors and test failures.
    
    \item Our empirical findings show that model accuracy declines significantly as \hunkdivergence increases and spatial proximity decreases. All evaluated models without augmentation failed to repair any bugs in the most dispersed \emph{Fragment} class, underscoring the challenges posed by highly scattered hunks.
\end{itemize}

% \ali{sounds like future work. Let's stick to what the paper does for now}
%Together, these contributions lay the groundwork for complexity-aware multi-hunk repair. These findings motivate new strategies that leverage predicted divergence categories to guide model selection, balancing accuracy and inference cost; context granularity, choosing minimal yet sufficient code regions; and selectively invoking higher-capacity models for high-divergence patches.
%\input{motivating-example}
\section{Characterization of Multi-hunk Patches}
\label{sec:characterization}

Although multi-hunk patches have been studied~\cite{hercules, iter, repair-agent}, their structural and spatial properties and how they affect automated repair remain poorly understood.

% \change
\begin{definition}[Hunk]\label{def:hunk} A \emph{hunk} is a block of consecutive edits applied to a specific region in the source code.
We define each hunk \( h_i \) as a tuple:
\[
h_i = (\text{loc}_i, \text{content}_i, \text{file}_i, \text{method}_i, \text{pkg}_i)
\]
% \[
% h_i = (\text{loc}_i, \text{content}_i, \text{file}_i, \text{method}_i)
% \]
where:
\begin{itemize}[noitemsep,leftmargin=1.5em]
  \item \( \text{loc}_i \): line range of the edit,
  \item \( \text{content}_i \): token sequence derived from the raw diff (concatenated deletions and additions),
  \item \( \text{file}_i \): file path of the edit,
  \item \( \text{method}_i \): enclosing method identifier, or \texttt{None} if outside any method. 
  \item \( \text{pkg}_i \): package path as a sequence of directory segments.
\end{itemize}
\end{definition}
% \stopchange{}

\begin{definition}[Multi-Hunk Patch]\label{def:multihunk-patch}
A \emph{multi-hunk patch} \( P \) is a set of \( n \geq 2 \) distinct hunks, denoted as \( P = \{h_1, h_2, \dots, h_n\} \), where the hunks modify non-contiguous regions of source code.  \( P \) is an atomic change intended to address a single bug.
\end{definition}

Our insight is that multi-hunk patches can vary significantly along several dimensions. In some cases, the hunks are nearly identical, such as repeated renamings or boilerplate edits, while in others, they involve semantically distinct changes that differ both lexically and structurally. These edits can also vary in proximity, from hunks located within the same method to those scattered across multiple files. This observation led us to ask: to what extent do hunks within a patch differ, and can this variation be meaningfully quantified? 

We present the first in-depth characterization of multi-hunk patches by introducing two new metrics: (a) \hunkdivergence, which quantifies lexical, structural, and file-level dissimilarity among hunks, and (b) \proximityclass, which classifies the spatial layout of hunks across the code hierarchy. 

\subsection{Hunk Divergence}
\label{sec:hunk-divergence}

%\keheliya{Alternative names for \textbf{hunk divergence}: Hunk Heterogeneity, Hunk Disparity} \ali{Dispersal focuses on location. Heterogeneity is good, but long and difficult to remember and say. I think we should keep hunk divergence since it measures the difference between pairs of hunks (quantitatively, with a distance metric). I also checked this with multiple LLMs, and they all suggest divergence as the best term to describe what we are measuring (location, content, structure).}

We propose \hunkdivergence, a new metric that captures lexical, structural, and file-level differences among hunks in a patch.

\begin{definition}[Pairwise Hunk Divergence]\label{def:pairwise-divergence}
We define the \emph{pairwise hunk divergence} between two hunks $h_i$ and $h_j$ of  \( P \)  through a combination of lexical, structural, and file-level separation distances:
\[
Div(h_1, h_2) = \frac{D_{\text{lex}} \cdot (D_{\text{ast}} + \gamma \cdot D_{\text{file}})}{1 + \gamma}
\]

where:
\begin{itemize}  
  \item $D_{\text{lex}}(h_i, h_j)$ quantifies the lexical distance between the code token sequences $\text{content}_i$ and $\text{content}_j$, where each token sequence is derived directly from the raw textual diff of the corresponding hunk. 
  
  \item $D_{\text{ast}}(h_i, h_j)$ measures the structural distance between the abstract syntax tree (AST) nodes of the two hunks.

  \item $D_{\text{file}}(h_i, h_j)$ quantifies the file-level separation between the files $\texttt{file}_i$ and $\texttt{file}_j$ that contain the hunks $h_i$ and $h_j$, respectively.
  
  \item $\gamma$ is a context-sensitive weighting factor that amplifies the contribution of inter-file separation to the overall divergence score, increasing the impact of file-based fragmentation when edits span distinct files in the repository.
  
\end{itemize}

\noindent
By construction, if $D_{\text{lex}}$, $D_{\text{ast}}$, and $D_{\text{file}}$ are each normalized to the interval $[0, 1]$, then the resulting pairwise divergence score satisfies $\text{Div}(h_i, h_j) \in [0, 1]$.
\end{definition}

%\header{Divergence Calculation}
To compute \emph{lexical distance ($D_{\text{lex}}$)}, we tokenize the diff exactly as it appears in the patch, preserving the ordering of added and deleted lines. For hunks involving both additions and deletions, we concatenate the deleted lines (before the change) and the added lines (after the change) prior to tokenization. In cases where a hunk contains only additions or only deletions, the token sequence consists solely of the respective lines as shown in the diff. We use BLEU~\cite{bleu} to quantify lexical similarity between token sequences of code fragments, following prior work~\cite{zhou:vulnerability-repair:icse24, shirafuji:code-repair:icast23, gu:deep-api-learning:fse16, jiang:commit-message-nmt:ase17, watson:assert-statement:icse20, glance, reptory, apr-survey:tosem24, koutcheme:distance-measure-repair:icer23}.

\[
D_{\text{lex}} = 1 - \textsc{BLEU}(T_1, T_2)
\]
where $T_1$ and $T_2$ are the token sequences of hunks $h_1$ and $h_2$, respectively.

%\keheliya{A hunk can span multiple lines and may not neatly map to a single AST node. How are we handling this case?}
The goal of \emph{structural distance} (\( D_{\text{ast}} \)) is to quantify how far apart two buggy hunks in the AST are when they appear in the same file. For this, we compute the shortest path between their corresponding nodes in the AST of the \emph{buggy} (pre-patch) version of the file. Each hunk is associated with the AST node corresponding to the first statement that begins within its line range. For deletions, we extract this node from the pre-patch AST using the first deleted line in the hunk. Let \( f = \text{file}(h_1) = \text{file}(h_2) \) be the shared file, and let \( T_f \) denote the buggy file’s AST. Let \( n_1, n_2 \in T_f \) be the AST nodes associated with \( h_1 \) and \( h_2 \), respectively. The structural distance is then defined as the normalized distance between these nodes, using the diameter of \( T_f \) as a scaling factor. 

To capture local structural variation between hunks, we normalize their AST distance using a natural logarithm-scaled formulation.
\[
D_{\text{ast}} =
\begin{cases}
\frac{\ln(1 + \textsc{ASTDist}(n_1, n_2))}{\ln(1 + \textsc{TreeDiameter}(\mathcal{T}_f))} & \text{if } \textsc{TreeDiameter}(\mathcal{T}_f) > 0 \\
0 & \text{otherwise}
\end{cases}
\]
\noindent
This formulation ensures numerical stability by avoiding division by zero when the AST is malformed or corresponds to an empty file. We apply log normalization to reduce the impact of large tree diameters, which can otherwise suppress meaningful structural differences between closely located nodes. If the buggy hunks occur in different files, we set $D_{\text{ast}} = 1$. This reflects that structural comparison is undefined across separate ASTs, and maximum structural divergence is used in this case.

% \header{Directory Divergence ($D_{\text{dir}}$)}
%\keheliya{Alternative names for \textbf{Directory Divergence}: Directory Distance, Directory Proximity}
% The goal of \emph{file divergence} ($D_{\text{file}}$) is to quantify the degree of file-level separation between two hunks within a patch. Let $\texttt{file}(h_i)$ denote the source file associated with hunk $h_i$. We define $D_{\text{file}} : \mathcal{H} \times \mathcal{H} \rightarrow \{0, 1\}$ as:

% \[
% D_{\text{file}}(h_i, h_j) =
% \begin{cases}
% 0 & \text{if } \texttt{file}(h_i) = \texttt{file}(h_j) \\
% 1 & \text{if } \texttt{file}(h_i) \neq \texttt{file}(h_j)
% \end{cases}
% \]

% This binary formulation captures whether a pair of edits is confined to the same file or spans across distinct files, serving as a proxy for the inter-file coordination effort required during multi-hunk repair.

The goal of \emph{file divergence} ($D_{\text{file}}$) is to quantify the degree of file-level separation between two hunks based on the layout of their enclosing file paths within the repository hierarchy. Let $\texttt{file}(h_i)$ denote the fully qualified file path associated with hunk $h_i$, represented as an ordered sequence of directory segments. For brevity, we write $f_i = \texttt{file}(h_i)$ and define $D_{\text{file}} : \mathcal{H} \times \mathcal{H} \rightarrow [0, 1]$ as:

\[
\begin{aligned}
D_{\text{file}}(h_i, h_j) =
\begin{cases}
0 & \text{if } f_i = f_j \\
1 - \dfrac{|\texttt{LCP}(f_i, f_j)|}{\max\big(|f_i|,\ |f_j|\big)} & \text{otherwise}
\end{cases}
\end{aligned}
\]
\noindent
% \ali{what are a and b? have we defined them already somewhere?}
Here, $\texttt{LCP}(f_i, f_j)$ denotes the length of the longest common prefix between the directory sequences $f_i$ and $f_j$, the two distinct buggy files, and $|\cdot|$ denotes the number of directory components in a given path. This formulation produces a normalized estimate of file-level distance: file paths that share a deeper hierarchical structure yield lower divergence values. This provides a continuous, repository hierarchy-aware quantification of inter-file separation that is both bounded and independent of path depth. It captures whether a pair of hunks is confined to the same file or spans across distinct files, serving as a proxy for the inter-file coordination effort required during multi-hunk repair.

% \keheliya{should we explicitly say that once hunks are in different files, the formula reduces to $Div(h_i, h_j) = D_lex(h_i, h_j)$? I think it will help the reader to understand our intention more easily.}

Overall, our formulation of pairwise divergence $Div(h_1, h_2)$ ensures that the score remains bounded in $[0, 1]$, while amplifying file-level separation when edits span multiple files. For same-file edits ($D_{\text{file}} = 0$), the score is proportionally downweighted by $1 + \gamma$, reflecting the reduced coordination effort relative to inter-file changes.

\begin{algorithm}[t]
\SetAlgoLined
\DontPrintSemicolon
\caption{\textsc{Pairwise Divergence Score}}
\label{alg:pairwise-divergence}
\footnotesize
\KwIn{Hunks $h_1$, $h_2$ with attributes: \texttt{file}, \texttt{ast}, \texttt{code};\\
}
\KwOut{Normalized divergence score $D(h_1, h_2) \in [0, 1]$}

$\gamma \leftarrow 1$ \tcp*{Scaling factor for file-level divergence} \label{alg1-gamma}

$D_{\text{lex}} \leftarrow 1 - \textsc{BLEU}(h_1.\texttt{code},\ h_2.\texttt{code})$\; \label{alg1-lexical}

\uIf{$h_1.\texttt{file} = h_2.\texttt{file}$}{ \label{alg1-structural}
  $f \leftarrow h_1.\texttt{file}$\;
  $\mathcal{T}_f \leftarrow \textsc{ParseAST}(f)$\;
  $treeDiameter \leftarrow \textsc{ComputeDiameter}(\mathcal{T}_f)$\;
  
  \uIf{$treeDiameter > 0$}{
    $n_1 \leftarrow \textsc{LocateNode}(\mathcal{T}_f,\ h_1.\texttt{loc})$\;
  
    $n_2 \leftarrow \textsc{LocateNode}(\mathcal{T}_f,\ h_2.\texttt{loc})$\;
  
    $astDist \leftarrow \textsc{ASTDistance}(n_1,\ n_2)$\;
  
    $D_{\text{ast}} \leftarrow \ln(1 + astDist) / \ln(1 + treeDiameter)$\; \label{alg1-ast-normalization}
  }
  \Else{
    $D_{\text{ast}} \leftarrow 0$ 
  } \label{alg1-normalization}
  $D_{\text{file}} \leftarrow 0$ \tcp*{No file-level separation: edits are in the same file}
} \label{alg1-intrafile}
\Else{ 
  $D_{\text{ast}} \leftarrow 1$ \tcp*{Maximum divergence across separate ASTs} \label{alg1-crossfile}
  
  $f_1 \leftarrow h_1.\texttt{file}$,\quad $f_2 \leftarrow h_2.\texttt{file}$\;

  $lcpLen \leftarrow \textsc{LCP}(f_1,\ f_2)$\;

  $D_{\text{file}} \leftarrow 1 - lcpLen / \max(|f_1|,\ |f_2|)$\;

  $\gamma \leftarrow 2$ \tcp*{Amplify inter-file separation weight}
}

$pairDiv \leftarrow D_{\text{lex}} \cdot (D_{\text{ast}} + \gamma \cdot D_{\text{file}})$\;

$maxDiv \leftarrow 1 + \gamma$\;

\Return $D(h_1, h_2) \leftarrow pairDiv / maxDiv$\;
\end{algorithm}

In Algorithm~\ref{alg:pairwise-divergence}, we show the steps for computing the \emph{pairwise divergence $D(h_1, h_2)$} between two hunks. Line~\ref{alg1-lexical} computes lexical divergence using \textsc{BLEU} score.
%, a symmetric token-level metric that captures surface-level similarity while being robust to minor reordering. 
Lines~\ref{alg1-structural}--\ref{alg1-intrafile} implement the conditional logic for determining whether the hunks are co-located in the same file. Line~\ref{alg1-ast-normalization} handles the intra-file case: the structural divergence is computed as the normalized distance between AST nodes within the same file, using the AST’s diameter for normalization. File-level divergence is omitted by setting $D_{\text{file}} = 0$.
% and the weighting factor $\gamma$ is set to 1.0 to reflect the constrained scope. 
In contrast, the cross-file case (beginning at Line~\ref{alg1-crossfile}) assigns $D_{\text{ast}} = 1$.0 to denote maximal syntactic divergence across disjoint ASTs.
File-level divergence is then computed using the normalized longest common prefix (LCP) between the two file paths. 
% Specifically, $D_{\text{file}} = 1 - \frac{|\texttt{LCP}(f_1, f_2)|}{\max(|f_1|,\ |f_2|)}$, where $f_1$ and $f_2$ are the fully qualified file paths of $h_1$ and $h_2$, respectively. 
The scaling factor $\gamma$ is set to $2.0$ to amplify the influence of file-level separation. This dynamic adjustment of $\gamma$ reflects the increased effort required when developers must coordinate edits across multiple files, ensuring that file-level fragmentation contributes more significantly to the overall divergence score in multi-hunk patches.

%\begin{definition}[Hunk Divergence]\label{def:divergence}
%The \emph{hunk divergence} score of a multi-hunk patch quantifies the internal fragmentation of its hunks (Definition~\ref{def:multihunk-patch}) by aggregating the divergence across its constituent hunks. It is computed as the average pairwise divergence over its unordered hunk pairs:
%\[
%\text{Div}(P) = \frac{2}{n(n-1)} \sum_{1 \leq i < j \leq n} \text{Div}(h_i, h_j).
%\]
%\end{definition}

% \begin{definition}[Hunk Divergence]\label{def:divergence}
% The \emph{hunk divergence} of a multi-hunk patch \( P \) quantifies the degree of internal variation among its constituent hunks. It captures how lexically, structurally, and spatially dissimilar the hunks are from one another. The divergence score is computed as the average pairwise divergence over all unordered hunk pairs:
% \[
% \text{Div}(P) = \frac{2}{n(n-1)} \sum_{1 \leq i < j \leq n} \text{Div}(h_i, h_j),
% \]
% \end{definition}

\begin{definition}[Hunk Divergence]\label{def:divergence}
The \emph{hunk divergence} of a multi-hunk patch \( P \) quantifies the degree of internal variation among its constituent hunks. Let \( P \) consist of \( n \) hunks \( \{h_1, h_2, \dots, h_n\} \), we define the hunk divergence of \( P \) as:
\[
\text{Div}(P) = \ln(n) \cdot \left( \frac{2}{n(n-1)} \sum_{1 \leq i < j \leq n} \text{Div}(h_i, h_j) \right)
\]
\noindent
$\text{Div}(P)$ captures how lexically, structurally, and spatially dissimilar the hunks are pairwise and the coordination complexity introduced by the number of hunks in $P$.
\end{definition}

%\ali{note that we are using ln not log in the calculation. They do give different outcomes.}
% The result we reported is with log, we will update it to ln. Also updated the formula for AST normalization to ln. We will also change it make it consistent.

%ali: ln(5) = 1.60 vs log(5) = 0.69

% We just computed divergence using the current formula with a log(n) factor based on hunk count, and observed that Sprawl now appears more divergent than Fragment:
% 
% Nucleus,0.3176
% Cluster,0.3358
% Orbit,0.9178
% Sprawl,1.0957
% Fragment,0.9745
%
% This outcome likely results from the metric not accounting for the number of files modified by a patch. 
% 
% Our current hunk divergence metric accounts for lexical/structural/hunks variation, but not for whether the hunks span multiple files. Since multi-file patches likely introduce additional coordination complexity, should we consider extending the metric with a file-aware factor? For example:
%
% Div(P) = ln(n) * (2 / (n(n-1)) * sum_{i < j} Div(h_i, h_j)) * ln(f + 1)
%
% where f is the number of unique files.

% Ali: yes, that was kind of bothering me as well. Good catch and idea to extend the metric with a file-aware factor. Although then do we also need to make it method-aware?!! Maybe do file-aware first and see how that impacts the relation between the spatial proximity classes and divergence score.
%
%
% Nashid: definitely, we will collect the divergence score for file-aware factor first.

We apply a natural logarithmic scaling factor $\ln(n)$ to reflect coordination complexity, increasing divergence for more fragmented patches while avoiding over-penalization when hunks are highly similar. As a result, the overall hunk divergence score satisfies \( \text{Div}(P) \in [0, \ln(n)] \).

Consider the following examples: for a patch with \( n = 5 \) hunks that are all \emph{identical}, all pairwise distances are zero, yielding a hunk divergence score of \( \text{Div}(P) = 0 \). At the other extreme, if all five hunks are \emph{maximally different} (i.e., each pair has a divergence score of 1), then the average pairwise divergence is 1 and the resulting hunk divergence is \( \text{Div}(P) = \ln(5) \approx 1.61 \). In a more typical case, if the average pairwise divergence among the hunks is moderate (e.g., 0.5), the divergence score becomes \( \text{Div}(P) = \ln(5) \cdot 0.5 \approx 0.80 \). %These examples highlight how the metric reflects both the internal heterogeneity of edits and the coordination overhead introduced by hunk count.

% \header{Hunk Count}
% \ali{I mentioned this in our last meeting, but the more I think about it the more important it becomes:} 

% The problem: two patches could have the same average pairwise divergence, but one may have 10 hunks and another 3 hunks—the one with 10 hunks is almost certainly harder to repair. Our current definition of Hunk Divergence accounts for average dissimilarity, but it treats all patches with 3 vs. 10 dissimilar hunks the same if their pairwise averages match.

% Possible Solutions:

% We could introduce a variant metric (or a normalized augmentation) that multiplies or scales the average pairwise divergence by a function of hunk count:

% \begin{definition}[Cardinality-Weighted Hunk Divergence]\label{def:weighted-divergence}
% Let \( P \) be a patch with \( n \) hunks. The cardinality-weighted hunk divergence adjusts the average pairwise divergence to reflect the number of hunks:
% \[
% \text{Div}_w(P) = \log(n) \cdot \text{Div}(P)
% \]
% \end{definition}

% The logarithmic factor avoids over-penalizing patches with a high number of similar hunks.

% We can choose a different function (e.g., n, $\sqrt{n}$, etc.) depending on empirical results or intuitions.

% We could also include this in the Hunk Divergence definition (\ref{def:divergence}) to avoid having more definitions and metrics.

\subsection{Spatial Proximity}
% While \emph{hunk divergence} provides a fine-grained, quantitative measure of distance between edits, it does not capture the broader spatial layout of the code in which the developer is making changes. To complement this, we introduce \emph{spatial proximity}, a rule-based classification that categorizes multi-hunk patches based on the dispersion of their edits across methods, files, and packages in the codebase.

%\emph{Hunk divergence} provides a fine-grained measure of distance between hunks of a patch. This continuous formulation enables quantitative analysis of hunk heterogeneity. To complement this, we introduce \emph{spatial proximity}, a rule-based classification that categorizes multi-hunk patches based on the dispersion of their edits across methods, files, and packages in the codebase.

% renamed Scattershot -> Fragment
\emph{Spatial proximity} classifies each multi-hunk patch into one of five categories---\emph{Nucleus}, \emph{Cluster}, \emph{Orbit}, \emph{Sprawl}, or \emph{Fragment}---representing increasing levels of proximity dispersion across methods, files, and packages. Unlike hunk divergence, which assigns continuous quantitative scores based on hunk dissimilarity, this classification provides rule-based insights into the spatial distribution of hunks.

\begin{definition}[Spatial Proximity Class]\label{def:spatial-proximity-class}
Let \( P = \{h_1, \dots, h_n\} \) be a multi-hunk patch. 
% Let \( P = \{h_1, \dots, h_n\} \) be a multi-hunk patch. For each hunk \( h_i \in P \), let \( m_i = \texttt{method}(h_i) \), \( f_i = \texttt{file}(h_i) \), and \( \mathit{pkg}_i = \texttt{package}(h_i) \), denoting its enclosing method, file, and package path, respectively. We define the following predicates based on the attributes of each hunk \( h_i \) (Definition~\ref{def:hunk}):

% \change
\begin{align*}
\mathrm{SM}(P)  &= \exists m \; \text{such that} \; \forall h_i \in P,\; m_i = m \\
\mathrm{SF}(P)  &= \exists f \; \text{such that} \; \forall h_i \in P,\; f_i = f \\
% \mathrm{SP}(P)  &= \exists \mathcal{P} \; \text{such that} \; \forall h_i \in P,\; \mathcal{P}_i = \mathcal{P} \\
\mathrm{SP}(P) &= \exists p \; \text{such that} \; \forall h_i \in P,\; \text{pkg}_i = p \\
% \mathrm{LCP}(P) &= \min_{i \ne j} \left| \text{LCP}(\mathcal{P}_i, \mathcal{P}_j) \right| \\
\mathrm{LCP}(P) &= \min_{i \ne j} \left| \text{LCP}(\text{pkg}_i, \text{pkg}_j) \right| 
\end{align*}
% \stopchange{}

% \begin{align*}
% \mathrm{SM}(P)  &= \exists m \; \text{such that} \; \forall h_i \in P,\; m_i = m \\
% \mathrm{SF}(P)  &= \exists f \; \text{such that} \; \forall h_i \in P,\; f_i = f \\
% \mathrm{SP}(P)  &= \exists \mathit{pkg} \; \text{such that} \; \forall h_i \in P,\; \mathit{pkg}_i = \mathit{pkg} \\
% \mathrm{LCP}(P) &= \min_{i \ne j} \left| \text{LCP}(\mathit{pkg}_i, \mathit{pkg}_j) \right|
% \end{align*}

These predicates assess different aspects of spatial proximity among the edits in a patch. Specifically:

\begin{itemize}[leftmargin=1.5em]
  \item \( \mathrm{SM}(P) \): all hunks are in the same method.  
  \item \( \mathrm{SF}(P) \): all hunks are in the same file.  
  \item \( \mathrm{SP}(P) \): all hunks share the same package.
  \item \( \mathrm{LCP}(P) \): minimum shared directory depth across hunk pairs.
\end{itemize}

Then, the spatial proximity class C(P) is defined as:

% \begin{align*}
% \mathcal{C}(P) =
% \begin{cases}
% \texttt{Nucleus}       & \text{if } \mathrm{SF}(P) \land \mathrm{SM}(P) \\
% \texttt{Cluster}       & \text{if } \mathrm{SF}(P) \land \lnot \mathrm{SM}(P) \\
% \texttt{Orbit}         & \text{if } \lnot \mathrm{SF}(P) \land \mathrm{SP}(P) \\
% \texttt{Sprawl}        & \text{if } \lnot \mathrm{SF}(P) \land \lnot \mathrm{SP}(P) \land \mathrm{LCP}(P) > 3 \\
% \texttt{Fragment}   & \text{if } \lnot \mathrm{SF}(P) \land \lnot \mathrm{SP}(P) \land \mathrm{LCP}(P) \leq 3 \\
% \end{cases}
% \end{align*}
% \end{definition}

\begin{align*}
\mathcal{C}(P) =
\begin{cases}
\texttt{Nucleus}       & \text{if } \mathrm{SF}(P) \land \mathrm{SM}(P) \\
\texttt{Cluster}       & \text{if } \mathrm{SF}(P) \land \lnot \mathrm{SM}(P) \\
\texttt{Orbit}         & \text{if } \lnot \mathrm{SF}(P) \land \mathrm{SP}(P) \\
\texttt{Sprawl}        & \text{if } \lnot \mathrm{SF}(P) \land \lnot \mathrm{SP}(P) \land \mathrm{LCP}(P) > \lambda \\
\texttt{Fragment}      & \text{if } \lnot \mathrm{SF}(P) \land \lnot \mathrm{SP}(P) \land \mathrm{LCP}(P) \leq \lambda \\
\end{cases}
\end{align*}
\end{definition}

Here, \( \lambda \) is a threshold on the directory depth used to distinguish structurally dispersed patches (\textsc{Sprawl}) from widely scattered ones (\textsc{Fragment}). This distinction is motivated by the intuition that when edits occur in directory paths with minimal shared hierarchy, the coordination burden for developers increases substantially.
 
\subsection{\benchmarkdataset: A Dataset of Multi-Hunk Bugs}
To enable principled evaluations, we introduce \benchmarkdataset, a curated dataset of multi-hunk bugs derived from \textsc{Defects4J}~\cite{defects4j}. \textsc{Defects4J} includes reproducible, test-suite backed bugs from open-source Java projects across diverse domains. We analyze all the 835 bugs in \textsc{Defects4J}\footnote{We use \textsc{Defects4J} version 2.0.1 for our analysis.} and identify those whose developer-written patches contain multi-hunk fixes (Definition~\ref{def:multihunk-patch}). We found that 372 out of 835 bugs (44.6\%) are multi-hunk. These 372 multi-hunk bugs form the basis of our \benchmarkdataset for further analysis. Beyond extracting patch-level characteristics, we augment \benchmarkdataset with natural language context for each bug. Specifically, we collect bug reports, titles, and summaries from official project repositories and issue trackers, including Apache JIRA, GitHub Issues, and SourceForge. This enriched metadata facilitates prompt construction and retrieval-augmented strategies for LLM-based program repair.

By combining patches validated by test suites with detailed hunk information (Definitions~\ref{def:hunk}, \ref{def:multihunk-patch}), \benchmarkdataset enables analysis of heterogeneity across lexical, structural, file-level divergence, and spatial distribution.

\subsection{Hunk Characterization}
To analyze the complexity of multi-hunk bugs, we examine each patch in \benchmarkdataset. % in terms of its number of change locations (\emph{hunks}) and the distribution of these changes across files.
%
%\header{Coarse-Grained Characterization} 
Each patch is systematically analyzed based on the developer-written ground truth patch. We first identify the set of modified files by locating lines that begin with \code{diff --git}, which mark the inclusion of a file in the patch. Then, we extract distinct hunks within each file by detecting lines starting with \code{@@}, which indicate hunk boundaries. For single-hunk files, we further quantify the number of modified lines by counting those prefixed with either \code{-} (deletions) or \code{+} (additions). If a deletion is directly followed by an addition, we treat it as a single-line substitution rather than two independent changes. These steps facilitate a precise classification of hunks. 

\begin{table}%[htbp]
\scriptsize
\caption{Multi-hunk patch file scope and hunk count.}
\label{tab:grouped_bug_categories}
\centering
\begin{tabular}{@{}c|c|c@{}}
\toprule
\textbf{File Scope}       & \textbf{Hunk Count Category}         & \textbf{Number of Bugs} \\ \midrule
\multirow{3}{*}{Single-file} & Two hunks                     & 140                     \\
                             & Three hunks                   & 55                      \\
                             & Four or more hunks            & 49                      \\ \cmidrule(lr){1-3}
\multirow{3}{*}{Multi-file}  & Two hunks                     & 37                      \\
                             & Three hunks                   & 23                      \\
                             & Four or more hunks            & 68                      \\ \midrule
\multicolumn{2}{c|}{\textbf{Total}} & \textbf{372}            \\ \bottomrule
\end{tabular}
\end{table}

We categorize multi-hunk bugs along two axes: the number of change regions (\emph{hunks}) and the number of files touched in Table~\ref{tab:grouped_bug_categories}. Single-file multi-hunk bugs dominate, with 244 cases where changes remain confined within a single file—this includes 140 two-hunks, 55 with three hunks, and 49 with four or more. These reflect intra-file repair scenarios, requiring local reasoning. In contrast, 128 bugs span multiple files, representing more fragmented, distributed repairs. Notably, 68 of these involve four or more hunks, indicating considerable dispersion across the codebase. 
% For instance, \emph{multi-hunk patch} \( P \) for \textsc{Jsoup\_56}\footnote{\url{https://github.com/rjust/defects4j/blob/master/framework/projects/Jsoup/patches/56.src.patch}} touches 10 hunks across 5 files.
For instance, \emph{multi-hunk patch} \( P \) for \textsc{Jsoup\_56}~\cite{jsoup56-multi-hunk-patch} touches 10 hunks across 5 files.

%\ali{For the coarse-grain analysis, do we have any stats on how many files the multi-hunk multi-file bugs span? What are the min, max, mean, median, for example?} 

To complement our earlier categorization, Table~\ref{tab:multi_hunk_distribution} presents project-specific statistics capturing how multi-hunk bugs manifest across different codebases. In 15 out of 17 projects, the median number of hunks per bug is at most 3, suggesting that the majority of bugs involve relatively localized changes. 
% However, outliers exist, \texttt{Jsoup\_87}\footnote{\url{https://github.com/rjust/defects4j/blob/master/framework/projects/Jsoup/patches/87.src.patch}} has 47 hunks.
However, outliers exist, \texttt{Jsoup\_87}~\cite{jsoup87-multi-hunk-patch} has 47 hunks.
% In contrast, \texttt{Chart\_4}\footnote{\url{https://github.com/rjust/defects4j/blob/master/framework/projects/Chart/patches/4.src.patch}} involves a multi-hunk fix with only two hunks confined to a single method, which is simpler to reason about.
In contrast, \texttt{Chart\_4}~\cite{chart4-multi-hunk-patch} involves a multi-hunk fix with only two hunks confined to a single method, which is simpler to reason about.
%
% \begin{quote}
% \scriptsize
% \begin{verbatim}
% @@ -4490,7 +4490,6 @@
% -                if (r != null) {
%                      Collection c = r.getAnnotations();
%                      Iterator i = c.iterator();
%                      while (i.hasNext()) {
% @@ -4499,7 +4498,6 @@
%                              includedAnnotations.add(a);
%                          }
%                      }
% -                }
% \end{verbatim}
% \end{quote}
%
Across all 17 projects, the median number of files modified per bug does not exceed 2; nonetheless, notable exceptions are present. 
% For instance, \texttt{JacksonDatabind\_103}\footnote{\url{https://github.com/rjust/defects4j/blob/master/framework/projects/JacksonDatabind/patches/103.src.patch}} spans 26 hunks across 16 files, modifying exception handling across the codebase. 
For instance, \texttt{JacksonDatabind\_103}~\cite{jackson103-multi-hunk-patch} spans 26 hunks across 16 files, modifying exception handling across the codebase. Conversely, certain projects such as Lang, Collections, and JacksonXml contain no multi-file bugs.

%\ali{this part needs updating, I think, since the classification has changed}

\begin{table}[t]
\caption{\benchmarkdataset dataset: hunk and file statistics}
\label{tab:multi_hunk_distribution}
\centering
\scriptsize
\resizebox{\columnwidth}{!}{%
\begin{tabular}{l|r|rrrr|rrrr}
\toprule
\textbf{Project} & \textbf{Bugs} &
\multicolumn{4}{c|}{\textbf{Hunks per Bug}} &
\multicolumn{4}{c}{\textbf{Files per Bug}} \\
& &
\textbf{Min} & \textbf{Median} & \textbf{Mean} & \textbf{Max} &
\textbf{Min} & \textbf{Median} & \textbf{Mean} & \textbf{Max} \\
\midrule
Chart           & 10  & 2 & 2.50 & 3.00 & 6   & 1 & 1.00 & 1.20 & 2 \\
Cli             & 17  & 2 & 2.00 & 3.47 & 10  & 1 & 1.00 & 1.71 & 4 \\
Closure         & 82  & 2 & 3.00 & 4.06 & 21  & 1 & 1.00 & 1.63 & 6 \\
Codec           & 7   & 2 & 3.00 & 3.57 & 7   & 1 & 2.00 & 2.43 & 6 \\
Collections     & 1   & 3 & 3.00 & 3.00 & 3   & 1 & 1.00 & 1.00 & 1 \\
Compress        & 18  & 2 & 3.00 & 3.89 & 14  & 1 & 1.00 & 1.61 & 7 \\
Csv             & 4   & 2 & 2.00 & 2.50 & 4   & 1 & 1.00 & 1.25 & 2 \\
Gson            & 7   & 2 & 2.00 & 4.00 & 12  & 1 & 1.00 & 1.43 & 3 \\
JacksonCore     & 12  & 2 & 3.50 & 4.33 & 8   & 1 & 2.00 & 1.75 & 3 \\
JacksonDatabind & 55  & 2 & 3.00 & 4.58 & 26  & 1 & 1.00 & 1.87 & 16 \\
JacksonXml      & 3   & 2 & 2.00 & 2.67 & 4   & 1 & 1.00 & 1.00 & 1 \\
Jsoup           & 37  & 2 & 3.00 & 4.73 & 47  & 1 & 1.00 & 1.89 & 5 \\
JxPath          & 14  & 2 & 4.00 & 4.50 & 11  & 1 & 2.00 & 1.93 & 5 \\
Lang            & 25  & 2 & 2.00 & 2.72 & 7   & 1 & 1.00 & 1.00 & 1 \\
Math            & 50  & 2 & 2.00 & 2.80 & 12  & 1 & 1.00 & 1.26 & 7 \\
Mockito         & 20  & 2 & 3.00 & 4.00 & 20  & 1 & 1.00 & 1.40 & 5 \\
Time            & 10  & 2 & 3.00 & 4.00 & 10  & 1 & 1.00 & 1.50 & 3 \\
\midrule
\textbf{Total}  & \textbf{372} & \textbf{2} & \textbf{3.00} & \textbf{3.86} & \textbf{47} & \textbf{1} & \textbf{1.00} & \textbf{1.59} & \textbf{16} \\
\bottomrule
\end{tabular}
}
\end{table}

%\header{Fine-Grained Characterization}
%\ali{can we also analyze the indivisibility of the hunks as part of their characterization? How many of the bugs in Hunk4J are indivisible? which? We should definitely cite that paper here.}

%Beyond the count and distribution of hunks and files, we further characterize multi-hunk patches through their \emph{hunk divergence} (Definition~\ref{def:divergence}). 
Figure~\ref{fig:divergence-components} presents the distribution of divergence for \benchmarkdataset. 
Lexical distance ($D_{\text{lex}}$) (Md = 0.94, Mean = 0.82) is sharply skewed toward the upper bound, indicating that most hunk pairs are lexically dissimilar. Structural distance ($D_{\text{ast}}$) (Md = 0.6, Mean = 0.62) shows broader dispersion, reflecting moderate variation in the AST of edits.
% \ali{how about the file? say something about that too!}
% File-level distance for multi-files ($D_{\text{file}}$), by contrast, is heavily skewed toward zero (Md = 0.0, Mean = 0.09), indicating that the majority of multi-hunk patches remain within the same file.
% 
% now added for multi-file
File-level distance for multi-file patches ($D_{\text{file}}$) exhibits a moderate central tendency (Md = 0.250, Mean = 0.251), indicating that while many multi-hunk patches involve hunks within closely related files, a non-trivial portion span more distinct locations in the file hierarchy.

%While file divergence ($D_{\text{file}}$) is defined as a binary indicator between any two hunks, $0$ if the hunks belong to the same file, $1$ otherwise, the aggregate divergence over a multi-hunk patch is not binary. 
% All $\binom{4}{2} = 6$ hunk pairs:
% $(h_1, h_2)$ → same file (File A) → 0
% $(h_1, h_3)$ → different files → 1
% $(h_1, h_4)$ → different files → 1
% $(h_3, h_4)$ → same file (File B) → 0
% $(h_2, h_3)$ → different files → 1
% $(h_2, h_4)$ → different files → 1
% So you have:
% 2 intra-file pairs → contribute 0
% 4 inter-file pairs → contribute 1
%Consider a patch with four hunks: $h_1, h_2$ located in File A, and $h_3, h_4$ in File B. This yields $\binom{4}{2} = 6$ unordered hunk pairs. Among these, $(h_1, h_2)$ and $(h_3, h_4)$ are intra-file pairs; the remaining four involve edits across different files. File-level divergence is defined as the fraction of inter-file pairs among all hunk pairs. Thus, $D_{\text{file}} = \frac{4}{6} = 0.667$, reflecting a majority of edit pairs requiring cross-file coordination.

% \ali{mention the min and max as well}
% \ali{how is that substantial?}
% \daniel{checked the hunk divergence statistics for the min, max, and 75\% with python code for substantial analysis} 
Overall hunk divergence in \benchmarkdataset ranges from a minimum of 0.00 to a maximum of 1.60, with a median of 0.42, a mean of 0.47.
%, and a 75th percentile of 0.624.
% \keheliya{Cant parse next sentence. Is it needed?} 
% Rewrote the sentence, so closing this comment
% Such a wide spread—fully 25 \% of multi-hunk patches exceed 0.624 in divergence.
Notably, 25\% of multi-hunk patches exhibit divergence values greater than 0.624.
% This broad range indicates substantial variation in divergence, driven by differences in lexical content, code structure, and file-separation distance.
The distribution reflects heterogeneity across lexical content, code structure, and file-level separation. Importantly, the divergence landscape is far from trivial, refuting any assumption of homogeneity.

\begin{figure}[t]  
  \centering
  \resizebox{0.7\columnwidth}{!}{%
    \begin{tabular}{cc}
      \begin{subfigure}[b]{0.45\columnwidth}
        \includegraphics[width=\linewidth]{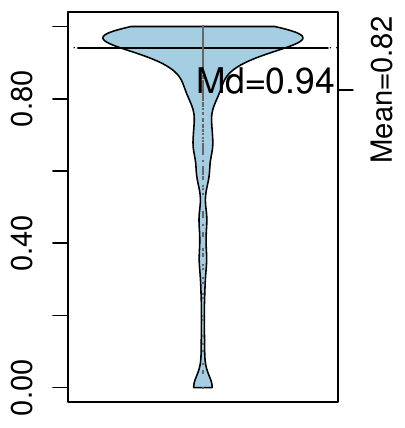}
        \caption{Lexical}
      \end{subfigure} &
      \begin{subfigure}[b]{0.45\columnwidth}
        \includegraphics[width=\linewidth]{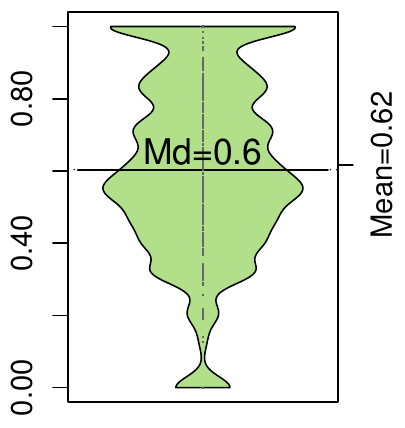}
        \caption{Structural}
      \end{subfigure} \\
      \begin{subfigure}[b]{0.45\columnwidth}
        \includegraphics[width=\linewidth]{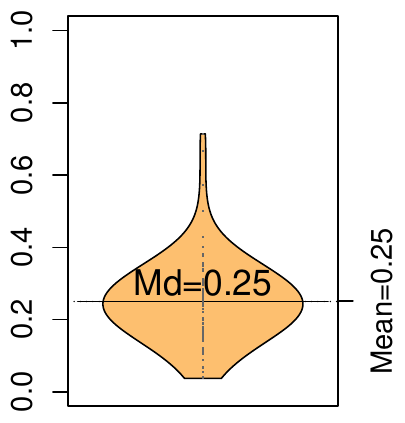}
        \caption{File}
      \end{subfigure} &
      \begin{subfigure}[b]{0.45\columnwidth}
        \includegraphics[width=\linewidth]{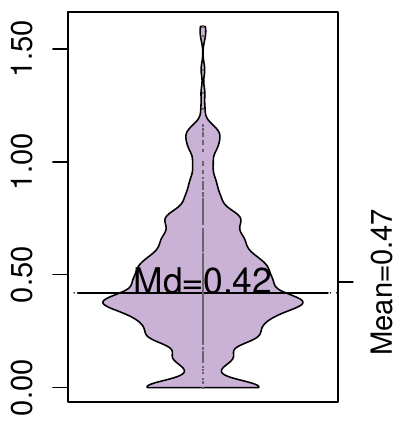}
        \caption{Overall}
      \end{subfigure}
    \end{tabular}
  }
  \caption{Distribution of different divergence components and the overall hunk divergence.}
  \label{fig:divergence-components}
\end{figure}

We also characterize the multi-hunk patches according to our spatial proximity classification (Definition~\ref{def:spatial-proximity-class}). We instantiate the proximity threshold \( \lambda \) based on the empirical distribution of package path depths in \benchmarkdataset. The median package depth across all buggy files is 6; we set \( \lambda = 3 \), corresponding to half this value, to conservatively separate loosely grouped edits from those that are structurally dispersed across the codebase. %Patches whose hunks fall within moderately shared directory structures are classified as \textsc{Sprawl}, while those spanning minimally overlapping paths are labeled \textsc{Fragment}. \textsc{Fragment} class exhibit the highest spatial dispersion, highlighting the coordination challenges posed by semantically distant edits. 

% \ali{first talk about the spatial proximity classification in the table before focusing on hunk divergence. mention some of the numbers in the clusters, is there anything interesting? for example Cluster is the largest with 185 followed by Orbit. Fragment is 11 which is also interesting...}
Table~\ref{tab:proximity_class} categorizes multi-hunk patches based on their spatial proximity, revealing a distinct skew toward more localized fixes. The \emph{Cluster} class, comprising 185 bugs, is the most prevalent and includes patches where multiple hunks occur within the same file but in different methods. This is followed by the \emph{Orbit} class (67 bugs), which captures cases where hunks span multiple files within the same package, and the \emph{Nucleus} class (59 bugs), where all hunks are confined to a single method. \texttt{Mockito\_6}~\cite{mockito6-multi-hunk-patch} from the \emph{Cluster} class contains 20 hunks, all confined to a single file yet distributed across multiple methods, demonstrating file-level cohesion despite method-level dispersion. In contrast, the \emph{Sprawl} (50 bugs) and \emph{Fragment} (11 bugs) classes represent increasingly dispersed patches. The \emph{Fragment} category consists of patches where hunks are scattered across unrelated packages. An example from \emph{Fragment} class is \texttt{Closure\_144}~\cite{closure144-multi-hunk-patch}, in which the hunks are scattered across unrelated packages, highlighting the low spatial locality. The average hunk divergence increases monotonically with spatial dispersion. \textsc{Fragment} patches, though rare, exhibit the highest divergence (0.7372), underscoring their complexity.

\begin{table}[t]
\scriptsize
\caption{Spatial proximity class and hunk divergence.}
\label{tab:proximity_class}
\centering
\begin{tabular}{l|c|c}
\toprule
\textbf{Proximity Class} & \textbf{Number of Bugs} & \textbf{Mean Hunk Divergence} \\
\midrule
Nucleus        & 59  & 0.2548 \\
Cluster        & 185 & 0.4280 \\
Orbit          & 67 & 0.5628 \\
Sprawl         & 50   & 0.6718 \\
Fragment       & 11   & 0.7372 \\ 
% Scattershot    & 0   & X.XX \\
% \midrule
\bottomrule
\end{tabular}
\end{table}

% !TEX root =  ../main.tex
\section{Empirical Evaluation}
\label{sec:evaluation}

Our study addresses the following research questions:

% \ali{we need to change our RQs. Here is what I propose:}

% Feedback from Prof
% - If you are fine with these new RQs, structure the evaluation section accordingly.
% - each RQ gets its own subsection and we only address that RQ there (no cross-cutting concerns)
% - BTW It is fine to include more information in the tables than needed for each RQ. However, only information directly related to the RQ should be discussed in each corresponding subsection.

\begin{itemize}

\item \textbf{RQ1}: How well do LLMs perform on multi-hunk repair tasks?

\item \textbf{RQ2}: To what extent do augmented prompting techniques (RAG, context, feedback) improve repair accuracy?

\item \textbf{RQ3}: How do hunk characteristics such as divergence and spatial proximity influence the repair success of LLM-based techniques?

\item \textbf{RQ4}: Are certain types of divergence and spatial proximity more predictive of challenges in LLM repair? 
%types of divergence: (lexical vs. structural vs. proximity)
%classes of spatial proximity: Nucleus, Cluster, Orbit, Sprawl, or Fragment

\end{itemize}

% \subsection{Benchmarking Infrastructure: \benchmark}
\subsection{Benchmarking Framework: \benchmark}
\label{subsec:birch}

To enable reproducible evaluations across different LLMs, we present \benchmark, a benchmarking framework for multi-hunk program repair. It offers an end-to-end pipeline, from input preparation to patch validation. The process begins by checking out the buggy version of the program from \benchmarkdataset and setting up a workspace. For each bug, \benchmark extracts the buggy code, relevant test cases, and associated metadata. It then constructs an input prompt that includes the buggy code, the precise bug location, a natural language description of the issue, and the failing test cases with their corresponding error messages. Once the prompt is prepared, it is submitted to a selected LLM. The patch is parsed and applied to the buggy code. The code is then compiled and tested using the test suite. Compilation errors and failing test cases are logged. 

\header{Modes}
\benchmark supports assessing vanilla LLMs but also provides various optional modes for enhancing LLMs. The \textit{retrieval-based example selection} mode supports both sparse and dense retrieval techniques. The \emph{enclosing context selection} mode extracts context at varying granularities, including the method, class, and file levels. 
The \emph{feedback loop} mode facilitates iterative patch refinement: if a generated patch fails to compile, the corresponding compilation error message is appended to the subsequent prompt. If the patch compiles but fails to pass the test suite, detailed test failure information is incorporated to guide further refinement attempts. If the code compiles and passes all tests, the patch is marked as successful. All evaluation artifacts, including prompts, generated code, diffs, and test logs, are persisted for inspection and analysis.

\header{Implementation}
\label{subsec:implementation}
\benchmark is implemented in Python and supports both proprietary models and open-source models via Ollama~\cite{ollama}. We use \texttt{litellm}~\cite{litellm} to interface with LLM APIs across providers. Prompts are templated using TOML, enabling integration of new prompt variants.
We employ \texttt{JavaParser} to parse ASTs, which are used to compute node-level distances between hunks, derive the tree diameter, and extract enclosing scopes such as methods and classes. For sparse retrieval, we use BM25; for dense retrieval, we employ \texttt{text-embedding-text-small-3} or \texttt{Mini-LM}, storing embeddings in a FAISS~\cite{johnson:billion-scale-search:ieeetran19} index for nearest‐neighbor lookup.

\subsection{Models}
% This study presents a comparative evaluation of 6 models, spanning both proprietary and open-source categories. Proprietary models include \textsc{gemini-2.5-flash} (1M token context window) developed by Google, which we evaluated in non-reasoning mode, \textsc{o4-mini} (200K tokens) and \textsc{gpt-4.1} (1M tokens) released by OpenAI, and \textsc{nova-pro} (300K tokens) accessed via Amazon Bedrock~\cite{amazon-bedrock}.  \emph{Open-source models} comprise Mistral AI's \textsc{mistral-large-2407} (128K tokens), and Meta's \textsc{LLaMA3-3-70b}, each supporting a 128K token context window. Except for OpenAI and Google offerings, all models were evaluated through Amazon Bedrock~\cite{amazon-bedrock} using full-parameter inference endpoints.
This study conducts a comparative evaluation of six models, selected to represent both proprietary and open-source variants with diverse architectural characteristics. The proprietary models include \textsc{Gemini-2.5-Flash} (1M token context window) from Google, evaluated in non-reasoning mode via the official Gemini API, and \textsc{Nova-Pro} (300K tokens) from Amazon. 
% From OpenAI, we evaluate both a reasoning model, \textsc{o4-mini} (200K tokens), and a non-reasoning variant, \textsc{GPT-4.1} (1M tokens). 
From OpenAI, we evaluate both a reasoning-capable model, \textsc{o4-mini} (200K tokens), and a non-reasoning variant, \textsc{GPT-4.1} (1M tokens), accessed directly via the OpenAI API. Parameter counts for proprietary models are not publicly disclosed.
The open-source models comprise \textsc{mistral-large-2407} (128K tokens, 123B parameters) from Mistral AI and \textsc{LLaMA3.3} (128K tokens, 70B parameters) from Meta. All models, except those from OpenAI and Google, were evaluated using full-parameter inference endpoints through Amazon Bedrock~\cite{amazon-bedrock}. Our goal is not to perform exhaustive model benchmarking. Rather, we aim to assess the generalizability of our findings across LLMs with varying architectures.

\subsection{Evaluation Metrics}
% \ali{are we the first to do this, or are other papers doing the same? if so cite those papers}
We assess patch correctness based on whether the generated code parses, compiles, and passes all tests provided by the benchmark suite. Following prior work~\cite{repairbench}, we report \emph{Plausible@1}, defined as the fraction of bugs for which the top-1 generated plausible patch passes the complete test suite:
\[
\text{Plausible@1} = \frac{1}{|\mathcal{B}|} 
    \sum_{b \in \mathcal{B}} 
    \mathrm{TestPass}(\mathsf{patch}_1^b)
\]
\noindent
where $\mathcal{B}$ is the set of bugs and $\texttt{patch}_1^b$ is the first generated patch for bug $b$. A patch is counted as plausible if it compiles and passes all test cases, including the originally failing one.

We generate one patch per bug using standard decoding settings (e.g., temperature = 0.0) for all models except \textsc{o4-mini}, whose API has a fixed temperature of 1.0.
% We do not sample multiple outputs. 
While decoding is configured to be deterministic at the API level, generation remains non-deterministic in practice due to low-level system factors such as thread scheduling, memory layout, and backend randomness~\cite{shuyin:non-determinism-of-chatgpt:tosem25}.

\begin{table}%[ht]
\scriptsize
\setlength{\tabcolsep}{4pt}
\centering
\caption{Multi-hunk bug repair with LLMs}
\label{tab:llm-pass1-eval}
\begin{tabular}{llc|cc|c|c}
\toprule
\textbf{Type} & \textbf{Model} & \textbf{Pass} & \textbf{Test Fail} & \textbf{Comp Fail} & \textbf{Plausible@1} & \textbf{Cost} \\
\midrule
\multirow{2}{*}{\makecell[l]{Open\\Source}} 
  & \texttt{Llama3.3}     & 46 & 73  & 253 & \textbf{12.37}\% & \$1.43 \\
  & \texttt{mistral-2407} & 45 & 106 & 221 & 12.10\% & \$4.08 \\
\midrule
\multirow{4}{*}{\makecell[l]{Propri-\\etary}} 
  & \texttt{nova-pro}     & 33 & 82  & 257 & 8.87\%  & \$2.30 \\  
  & \texttt{Gemini 2.5 Flash} & 50 & 32 & 290 & 13.44\% & \$0.37 \\
  & \texttt{GPT-4.1}      & 82 & 78  & 212 & 22.04\% & \$6.37 \\
  & \texttt{o4-mini}      & 100 & 57  & 215 & \textbf{26.88}\% & \$3.30 \\
\bottomrule
\end{tabular}
\end{table}

\subsection{Effectiveness of LLMs on Multi-Hunk Repair (RQ1)}

% \header{Proprietary Models Lead.}
% \ali{before saying who is best and worst, always describe the table and what it is presenting. Always.}
%Table~\ref{tab:llm-pass1-eval} presents results for 11 LLMs on multi-hunk repair, reporting \textsc{Plausible@1} accuracy, test and compilation failures, and total inference cost, grouped by open-source and proprietary models. The highest \textsc{Plausible@1} accuracy is achieved by \textsc{o4-mini} (26.88\%), followed by \textsc{GPT-4.1} (22.04\%) and \textsc{Gemini 2.5 Flash} (13.44\%). Open models such as \textsc{LLaMA3.3} trail behind (12.37\%), while smaller variants such as \textsc{Mistral-7B} register just above 1\%. 

%\ali{to me (as a reviewer) it does not make sense to highlight lower versions of models. We can have them in the tables but I don't think we should discuss them really.}
%\ali{performance or accuracy? performance in SE is about speed}
% \ali{Move cost discussions to the end of this subsection as that is not the main focus of the RQ.}
%Compilation failures dominate for smaller models, especially \textsc{LLaMA3-1-405B}, which fails to compile in $362$ of $372$ cases. These results show that despite promising advances, most LLMs struggle to produce correct multi-hunk repairs. 
%Proprietary models consistently outperform open-source alternatives, likely due to more extensive pretraining, larger context windows, and reinforcement learning from human feedback (RLHF).
% To evaluate the effectiveness of LLMs in repairing multi-hunk bugs, we fed all 372 bugs in our dataset through all 6 models.
To assess LLM effectiveness on multi-hunk repair, we ran all 372 bugs through the six models.
Table~\ref{tab:llm-pass1-eval} summarizes the results, including the number of fully repaired bugs (Pass), test failures, compilation failures, plausible patch rate (Plausible@1), and estimated cost per run. 
% Table~\ref{tab:llm-pass1-eval} reports repair outcomes, Plausible@1 rates, and cost per run.
% Overall, the best-performing model was \textsc{o4-mini}, which successfully repaired 100 of the 372 bugs (26.88\%), followed by \textsc{gpt-4.1} with 82 successful repairs (22.04\%). Among open-source models, the top performer was \textsc{LLaMA3.3}, repairing 46 bugs (12.37\%) followed by \textsc{mistral-2407} and \textsc{LLaMA3.2} each repair 45 bugs (12.10\%). In contrast, other open-source models such as \textsc{mistral-7b}, \textsc{mixtral-8x7b}, and \textsc{LLaMA3.1} performed substantially worse, with repair rates below 2\%.
Overall, the best-performing model was \textsc{o4-mini}, which successfully repaired 100 of the 372 bugs (26.88\%), followed by \textsc{gpt-4.1} with 82 successful repairs (22.04\%). Among open-source models, \textsc{LLaMA3.3} achieved the highest accuracy, repairing 46 bugs (12.37\%), followed closely by \textsc{mistral-2407}, which repaired 45 bugs (12.10\%).

Across the six LLMs, a total of $127$ unique bugs are resolved. %When considering the 7 open-source models collectively, they are able to generate plausible patches for $76$ unique bugs. In contrast, the four proprietary models, when combined, produce plausible patches for $122$ unique bugs. This highlights the superior collective coverage of proprietary models in addressing multi-hunk bugs, underscoring that open-source LLMs, while promising, have yet to reach parity in this domain.
%
% \ali{select to do what? unclear}
% To evaluate both their individual effectiveness and their collective potential in multi-hunk program repair, we select the best-performing LLM from each model family: \textsc{o4-mini} (OpenAI), \textsc{Gemini-2.5-Flash} (Google), \textsc{Nova-Pro} (Amazon), \textsc{Mistral-2407} (Mistral), and \textsc{LLaMA3-3-70B} (Meta). 
Among the two OpenAI models evaluated, we selected \textsc{O4-Mini} for further analysis due to its higher accuracy compared to \textsc{GPT-4.1}.
We then assess individual effectiveness and collective coverage using the following five models: \textsc{o4-mini}, \textsc{Gemini-2.5-Flash}, \textsc{Nova-Pro}, \textsc{Mistral-2407}, and \textsc{LLaMA3.3}.
Specifically, we conduct an intersection analysis on the sets of bugs each model successfully resolves. Together, these five models generate \textit{plausible} patches for $116$ unique bugs. 
% This finding underscores that substantial coverage in multi-hunk bug repair can be achieved through a strategically selected ensemble comprising the top model from each LLM family.
This finding underscores that substantial coverage in multi-hunk bug repair can be achieved through a carefully selected ensemble of models.
% We show the overlap of \textit{Plausible@1} patches generated by the LLM that performs the best for each model family in the Venn diagram shown in Figure~\ref{fig:venn-llm}. 
Figure~\ref{fig:venn-llm} presents a Venn diagram showing the overlap of \textit{Plausible@1} patches among the selected models.
Although all five models solve a shared pool of $16$ bugs, each contributes additional unique fixes, most notably \textsc{o4-mini}, which accounts for $39$ unique repaired bugs.

% When considering cost, model accuracy varies considerably with respect to inference cost. While \textsc{Gemini-2.5-Flash} achieves a Plausible@1 of $13.44\%$ at a total cost of only \$0.37, the lowest across all models, some open-source models such as \textsc{Mistral-2402} incur substantially higher costs (\$7.18) despite offering moderate accuracy. This disparity underscores that higher inference cost does not necessarily yield better repair accuracy.
When considering cost, model accuracy varies significantly relative to inference expense. \textsc{Gemini-2.5-Flash} achieves a Plausible@1 of $13.44\%$ at a total cost of just \$0.37, the lowest among all evaluated models. In contrast, \textsc{Mistral-Large-2407}, an open-source model, incurs a substantially higher cost of \$4.08 while yielding a comparable Plausible@1 of $12.10\%$. This disparity highlights that increased inference cost does not necessarily translate to higher repair accuracy.

Overall, despite recent advances, LLMs still struggle with multi-hunk bug repair, as even the best model fails to fix over 70\% of cases. The higher accuracy of \textsc{o4-mini} highlights the benefit of using reasoning-capable models with greater test-time compute compared to non-reasoning models. However, the limited repair accuracy even in the best-performing model suggests that increasing test-time compute alone may not be sufficient, particularly for complex repair tasks~\cite{snell:scaling-llm-test-time:iclr25}.

\begin{figure}%[ht]
  \centering
  \includegraphics[width=0.80\linewidth]{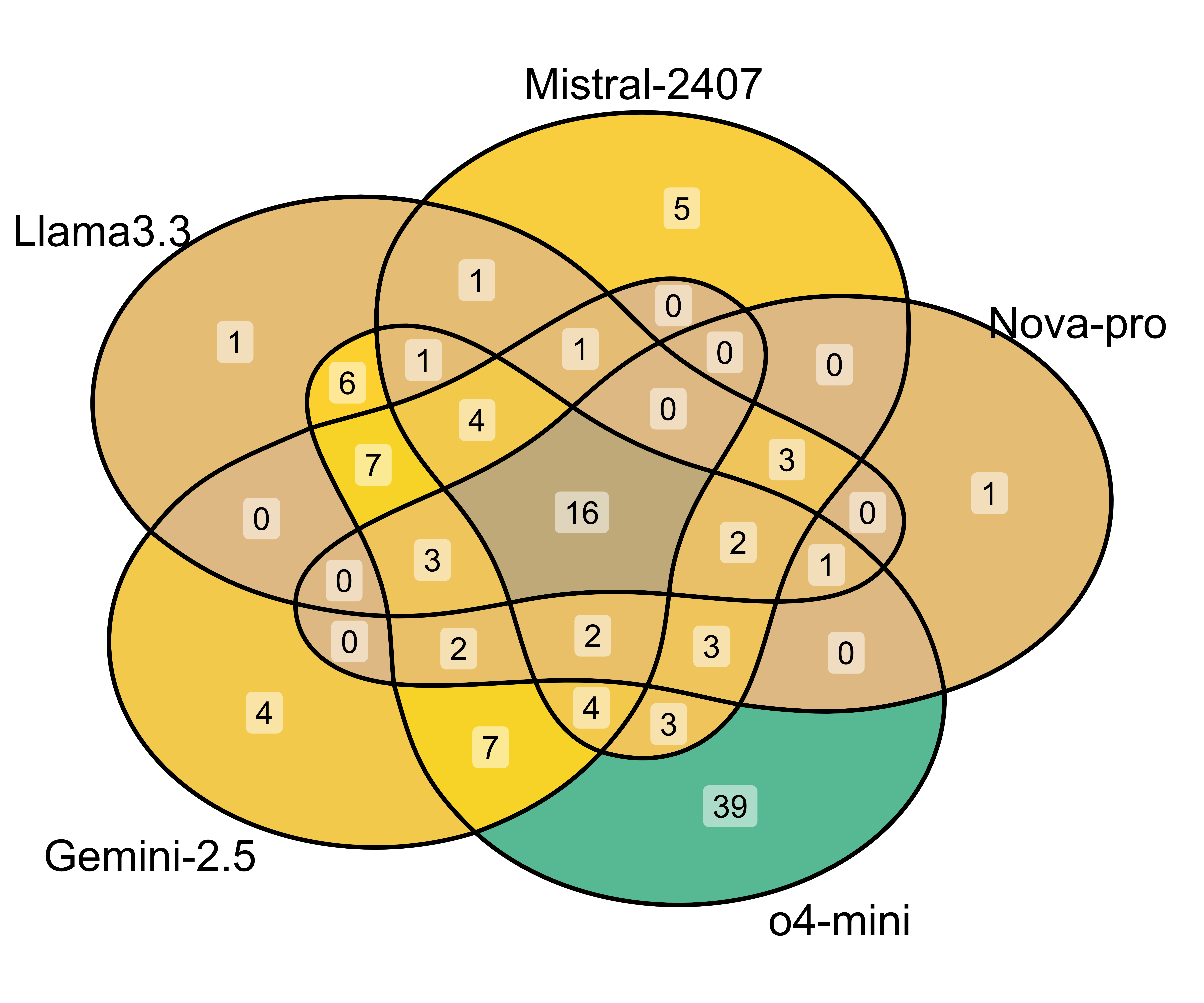}
  \caption{Overlap of fixes across model families}
  \label{fig:venn-llm}
\end{figure}

\subsection{Impact of Augmented Prompting (RQ2)}
\label{sec:augmented-prompting}
% \ali{somewhere in this subsection, we need to say that we only do this part for two LLMs, the best from open-source and proprietary}
% To assess how different prompting augmentations impact LLM-based program repair, we evaluate three aspects, (a) retrieval-augmented example selection, (b) contextual granularity, and (c) feedback-based refinement. Given the high inference cost of large-scale evaluation across all prompt augmentation strategies, we conduct our analysis on two representative models: \textsc{o4-mini}, the most accurate proprietary LLM, and \textsc{LLaMA3.3}, the best open-source model.
To assess the impact of different prompt augmentation strategies on LLM-based program repair, we evaluate three aspects: (a) retrieval-augmented example selection, (b) contextual granularity, and (c) feedback-based refinement. Given the high inference cost of large-scale evaluation across all prompt augmentation strategies, we limit this analysis to two representative models: \textsc{o4-mini}, the most accurate proprietary model, and \textsc{LLaMA3.3}, the most effective open-source model in our evaluation.

% \ali{where do the similar patches come from?}
We examine the effectiveness of \emph{retrieval-based} prompting by comparing different ways of representing and retrieving similar repair examples. Each buggy example is encoded either as code tokens or as AST elements, capturing lexical and structural perspectives, respectively. For each representation, we apply one sparse retrieval method and two dense embedding-based methods to identify semantically similar past patches as examples. By pairing the two code representations with these three retrieval techniques, we construct six example-selection strategies, each retrieving the top-5 most similar patches to include in the prompt. The example patches used in retrieval-augmented prompting are selected from multi-hunk patches in the \benchmarkdataset dataset that were fixed prior to the target bug---excluding the target bug itself---based on similarity in token or AST representations, depending on the retrieval configuration.

% Table~\ref{tab:rq2-prompt-augmentation} reports the effectiveness of various prompt augmentation strategies by presenting Plausible@1 accuracy, test failures, compilation failures, and inference cost for each configuration, evaluated on two representative models: \textsc{o4-mini} and \textsc{LLaMA3.3}. 
Table~\ref{tab:rq2-prompt-augmentation} reports the effectiveness of various prompt augmentation strategies in terms of accuracy, failure outcomes, and inference cost on \textsc{o4-mini} and \textsc{LLaMA3.3}. The results show that token-level representations paired with dense retrieval outperform both AST-based code views and the sparse retrieval (BM-25) technique. For \textsc{o4-mini}, the best configuration—token+\texttt{text-embed}—achieves a Plausible@1 of 31.18\%, surpassing the AST+\texttt{text-embed} variant (23.66\%) by 7.52 percentage points. Among retrieval strategies, \texttt{text-embed} consistently yields higher accuracy than MiniLM and BM25 across both token and AST inputs. Similar trends are observed for \textsc{LLaMA3.3}, where token+\texttt{text-embed} reaches 20.16\% Plausible@1, compared to 11.02\% with AST+\texttt{text-embed} and 17.20\% with token+BM25. This improvement can be attributed to how dense retrieval operates. Unlike sparse retrieval (BM25), which relies on token overlap and term frequency, dense retrieval (\texttt{text-embed}) encodes both the query and candidates into dense vectors using a neural embedding model. The embedding model was trained on token sequences, not structured ASTs. Consequently, similarity search is more accurate when both the query and candidates are represented as token sequences.

\begin{table}%[ht]
\scriptsize
\setlength{\tabcolsep}{3pt}
\centering
\caption{Multi-hunk bug repair with prompting strategies}
\label{tab:rq2-prompt-augmentation}
% \resizebox{\textwidth}{!}{%
\begin{tabular}{l l l c| c c| c | c}
\toprule
\textbf{Strategy} & \textbf{LLM} & \makecell{\textbf{Prompt}\\\textbf{Mode}}
  & \textbf{Pass} & \makecell{\textbf{Test}\\\textbf{Fail}}
  & \makecell{\textbf{Comp}\\\textbf{Fail}}
  & \makecell{\textbf{Plausible}\\\textbf{@1}} & \textbf{Cost} \\
\midrule
\multirow[c]{12}{*}{\shortstack[l]{RAG\\ (Example\\Selection)}}
 & \multirow[c]{6}{*}{\texttt{Llama3.3}}
   & AST + BM25          &  64 &  76 & 232 & 17.20\% & \$3.48 \\
 &                       & AST + MiniLM       &  42 &  96 & 234 & 11.29\% & \$9.09 \\
 &                       & AST + text-embed   &  41 &  94 & 237 & 11.02\% & \$7.94 \\
 &                       & Token + BM25       &  70 &  75 & 227 & 18.82\% & \$4.77 \\
 &                       & Token + MiniLM     &  69 &  75 & 228 & 18.55\% & \$3.75 \\
 &                       & Token + text-embed &  75 &  77 & 220 & \textbf{20.16}\% & \$4.00 \\
  \cmidrule(lr){2-8}
 & \multirow[c]{6}{*}{\texttt{o4-mini}}
   & AST + BM25          & 106 &  66 & 200 & 28.49\% & \$6.53 \\
 &                       & AST + MiniLM       &  91 &  74 & 207 & 24.46\% & \$14.90 \\
 &                       & AST + text-embed   &  88 &  79 & 205 & 23.66\% & \$13.25 \\
 &                       & Token + BM25       & 111 &  57 & 204 & 29.84\% & \$8.46 \\
 &                       & Token + MiniLM     & 114 &  57 & 201 & 30.65\% & \$8.38 \\
 &                       & Token + text-embed & 116 &  59 & 197 & \textbf{31.18}\% & \$7.26 \\
\midrule
\multirow[c]{6}{*}{\shortstack[l]{Context\\Granularity}}
 & \multirow[c]{3}{*}{\texttt{Llama3.3}}
   & Method             &  46 &  73 & 253 & 12.37\% & \$1.43 \\
 &                      & Class              &  43 & 113 & 216 & 11.56\% & \$3.80 \\
 &                      & File               &  58 & 132 & 182 & \textbf{15.59}\% & \$4.03 \\
  \cmidrule(lr){2-8}
 & \multirow[c]{3}{*}{\texttt{o4-mini}}
   & Method             & 100 &  57 & 215 & 26.88\% & \$3.30 \\
 &                      & Class              &  88 & 146 & 138 & 23.66\% & \$35.47 \\
 &                      & File               & 105 & 186 &  81 & \textbf{28.30}\% & \$43.35 \\
\midrule
\multirow[c]{2}{*}{\shortstack[l]{Feedback\\Loop}}
 & \texttt{Llama3.3}           & \shortstack[l]{Method + Code +\\text-embed} &  91 &  80 & 201 & \textbf{24.46}\% & \$7.64 \\  \cmidrule(lr){2-8}
 & \texttt{o4-mini}             & \shortstack[l]{Method + Code +\\text-embed} & 133 &  34 & 205 & \textbf{35.75}\% & \$14.63 \\
\bottomrule
\end{tabular}
% }
\end{table}

\begin{table*}[!ht]
\centering
\scriptsize
\caption{Hunk divergence for fixed and unfixed bugs, and spatial proximity distribution (\% fixed) across model families}
\label{tab:rq3-fixed-not-fixed}
\begin{tabular}{l|cccc|cccc|ccccc}
\toprule
\textbf{Model} 
  & \multicolumn{4}{c|}{\textbf{Hunk Divergence (Fixed)}} 
  & \multicolumn{4}{c|}{\textbf{Hunk Divergence (Unfixed)}}
  & \multicolumn{5}{c}{\textbf{Spatial Proximity (\% Fixed)}} \\
& Median & Mean & Max & Std Dev 
  & Median & Mean & Max & Std Dev 
  & Nucleus & Cluster & Orbit & Sprawl & Fragment \\
\midrule
\texttt{mistral-2407} 
  & 0.27 & 0.27 & 0.65 & 0.19 
  & 0.46 & 0.49 & 1.60 & 0.30 
  & 20.34 & 11.35 & 5.97  & 16.0 & 0.0 \\

\texttt{Llama3.3}    
  & 0.26 & 0.27 & 0.65 & 0.17 
  & 0.47 & 0.50 & 1.60 & 0.30 
  & 27.12 & 9.73  & 11.94 & 8.0  & 0.0 \\  

\texttt{nova-pro}     
  & 0.24 & 0.28 & 0.65 & 0.19 
  & 0.45 & 0.49 & 1.60 & 0.30 
  & 11.86 & 8.11  & 7.46  & 12.0 & 0.0 \\

\texttt{Gemini 2.5}   
  & 0.26 & 0.26 & 0.65 & 0.17 
  & 0.47 & 0.50 & 1.60 & 0.30 
  & 18.64 & 12.43 & 14.93 & 12.0 & 0.0 \\

\texttt{o4-mini}     
  & 0.29 & 0.28 & 0.70 & 0.18 
  & 0.50 & 0.53 & 1.60 & 0.30 
  & 49.15 & 23.78 & 23.88 & 22.0 & 0.0 \\
\bottomrule
\end{tabular}
% \vspace{-3.5ex}
\end{table*}

In addition to example selection, we study the effect of \emph{contextual granularity} by varying the scope of code included in the prompt. We compare three levels of enclosing context: method, class, and file. The results for enclosing context show that wider context generally improves repair success. For \textsc{o4-mini}, Plausible@1 increases from 26.88\% (method) to 28.30\% (file), with a drop at the class level (23.66\%), as class-level prompting often result in incomplete code generation. The reduced effectiveness at the class level can be attributed to LLMs interpreting the class snippet as an incomplete code fragment, often defaulting to patching only individual methods or lower-level constructs. In contrast, file-level prompts, which include import statements and top-level comments, provide clearer semantic cues, prompting the model to generate complete code. \textsc{LLaMA3.3} shows a similar pattern: file-level prompting yields 15.59\%, compared to 12.37\% (method) and 11.56\% (class). However, these gains come at a significant cost. For \textsc{o4-mini}, file-level prompts cost \$43.35, 13 times more than method-level (\$3.30), which achieves nearly comparable accuracy. 

% \begin{table*}%[ht]
% \centering
% \scriptsize
% \caption{Hunk Divergence and Spatial Proximity for Repaired Bugs (Across Model Families).}
% \label{tab:rq3-solved-only}
% \begin{tabular}{l|ccccc|c|ccccc}
% \toprule
% \textbf{Model} 
%   & \multicolumn{5}{c|}{\textbf{Hunk Divergence (Fixed)}} 
%   & \multicolumn{1}{c|}{\textbf{$p$-value}}
%   & \multicolumn{5}{c}{\textbf{Spatial Proximity (\% Fixed)}} \\
% & Min & Max & Mean & Median & Std Dev 
%   &
%   & Nucleus & Cluster & Orbit & Sprawl & Fragment \\
% \midrule
% \texttt{Mistral2407} 
%   & 0.0    & 0.65 & 0.27 & 0.27 & 0.19 
%   & \(3.30\times10^{-7}\)
%   & 20.34 & 11.35 & 5.97  & 16.0 & 0.0 \\

% \texttt{Llama3.3}    
%   & 0.0    & 0.65 & 0.27 & 0.26 & 0.17 
%   & \(6.88\times10^{-8}\)
%   & 27.12 & 9.73  & 11.94 & 8.0  & 0.0 \\  

% \texttt{nova-pro}     
%   & 0.0    & 0.65 & 0.28 & 0.24 & 0.19 
%   & \(2.79\times10^{-9}\)
%   & 11.86 & 8.11  & 7.46  & 12.0 & 0.0 \\

% \texttt{Gemini2.5}   
%   & 0.0    & 0.65 & 0.26 & 0.26 & 0.17 
%   & \(2.55\times10^{-5}\)
%   & 18.64 & 12.43 & 14.93 & 12.0 & 0.0 \\

% \texttt{o4-mini}     
%   & 0.0    & 0.70 & 0.28 & 0.29 & 0.18 
%   & \(3.16\times10^{-15}\)
%   & 49.15 & 23.78 & 23.88 & 22.0 & 0.0 \\
% \bottomrule
% \end{tabular}
% \end{table*}

% \keheliya{Shall we cite some SE+LLM papers here that use feedback from compiler/tests, self-debug?}
We further examine \emph{feedback-based refinement}, which augments prompting with dynamic execution information derived from compilation failures or test failures, following prior work~\cite{flakidock:icse25, ruan:specrover:icse25, bug-reprodcution}. To limit cost, the feedback prompt uses the method-level context and the best retrieval setup (token+\texttt{text-embed}). Despite this conservative design, feedback yields the highest accuracy across all settings. \textsc{o4-mini} achieves 35.75\% Plausible@1 and \textsc{LLaMA3.3} achieves 24.46\%, both surpassing their respective static prompt baselines. Notably, the accuracy of \textsc{o4-mini} with feedback surpasses that of the static prompt using file-level context ($28.30\%$), while costing less than half as much (\$14.63 vs. \$43.35). These results indicate that feedback, driven by runtime information from failed tests or execution traces, can guide the model more effectively than simply expanding the prompt context. Moreover, feedback mitigates the limitations of narrow context by incorporating dynamic execution traces, which provide concrete, failure-specific guidance that static prompts lack.

% \ali{What is the takeaway from these results for RQ2? Do augmentation prompting techniques solve the problem of multi-hunk bug repair? We need a punch line for each RQ.}
Although prompting augmentations such as feedback and retrieval improve repair accuracy over vanilla LLMs, even with the best configurations, 64\% of multi-hunk bugs still remain unfixed.

\subsection{Repair Outcomes by Hunk Divergence and Proximity (RQ3)}

\begin{table*}[ht]
\centering
\scriptsize
\caption{Hunk divergence for fixed and unfixed bugs, and spatial proximity (\% fixed) across best prompt strategies}
\label{tab:rq3-solved-not-solved-prompt-variants}
\resizebox{\textwidth}{!}{
\begin{tabular}{l l|cccc|cccc|ccccc}
\toprule
\textbf{Model} & \textbf{Technique Variant} 
  & \multicolumn{4}{c|}{\textbf{Divergence Score (Fixed)}}
  & \multicolumn{4}{c|}{\textbf{Divergence Score (Unfixed)}}
  & \multicolumn{5}{c}{\textbf{Spatial Proximity (\% Fixed)}} \\
  &
  & Median & Mean & Max & Std Dev 
  & Median & Mean & Max & Std Dev 
  & Nucleus & Cluster & Orbit & Sprawl & Fragment \\
\midrule
\multirow{3}{*}{\texttt{o4-mini}}
  & Token + text-embed 
    & 0.34 & 0.31 & 0.96 & 0.21 
    & 0.50 & 0.54 & 1.60 & 0.30 
    & 54.24 & 30.81 & 22.39 & 20.00 & 18.18 \\
  & File Context       
    & 0.39 & 0.40 & 1.08 & 0.23  
    & 0.46 & 0.49 & 1.60 & 0.31 
    & 37.29 & 33.51 & 13.43 & 22.00 & 9.09  \\
  & Feedback Loop      
    & 0.31 & 0.30 & 0.96 & 0.20 
    & 0.53 & 0.56 & 1.60 & 0.30 
    & 64.41 & 33.51 & 28.36 & 22.00 & 27.27 \\
\midrule
\multirow{3}{*}{\texttt{Llama3.3}}
  & Token + text-embed 
    & 0.28 & 0.28 & 0.78 & 0.21 
    & 0.47 & 0.52 & 1.60 & 0.29 
    & 33.90 & 18.92 & 17.91 & 14.00 & 9.09  \\
  & File Context       
    & 0.36 & 0.36 & 1.08 & 0.23  
    & 0.45 & 0.49 & 1.60 & 0.30 
    & 18.64 & 17.30 & 7.46  & 20.00 & 0.00  \\
  & Feedback Loop      
    & 0.30 & 0.29 & 0.96 & 0.20 
    & 0.49 & 0.52 & 1.60 & 0.30 
    & 42.37 & 21.62 & 25.37 & 14.00 & 18.18 \\
\bottomrule
\end{tabular}
}
\end{table*}
% \vspace{-4.0ex}

% To examine the conditions under which LLM-based program repair is most effective, we measure descriptive statistics of hunk divergence and the spatial class-wise accuracy percentages. These are derived from patches generated by the top-performing model in each of the five distinct LLM families.
To examine the conditions under which LLM-based program repair is most effective, we measure descriptive statistics of hunk divergence and spatial class-wise accuracy percentages. These metrics are derived from patches produced by the five models selected in RQ1, each representing a different model family.

Table~\ref{tab:rq3-fixed-not-fixed} reports the central tendencies of hunk divergence for both successfully and unsuccessfully fixed bugs. Additionally, it presents the proportion of bugs correctly fixed within each spatial proximity class.
% Table~\ref{tab:rq3-fixed-not-fixed} reports the central tendencies of hunk divergence for fixed and unfixed bugs, along with the proportion of correctly repaired bugs across spatial proximity classes. 

Successful repairs are consistently associated with low divergence across all five models (mean $\approx 0.26$–$0.28$; median  $\approx 0.24$–$0.29$). Variance remains narrow (standard deviation $< 0.20$), indicating high semantic alignment across generated patches. In contrast, unfixed bugs exhibit substantially higher hunk divergence (mean: $0.49$–$0.53$; median: $0.45$–$0.53$), along with broader variance and higher maxima (up to $1.60$). These results reflect the increased difficulty LLMs face in coordinating complex, semantically diverse edits across scattered code regions.

Table~\ref{tab:rq3-solved-not-solved-prompt-variants} reports our analysis across the best-performing prompting configurations, i.e., (1) \texttt{Token + text-embed}, (2) \texttt{File Context}, and (3) \texttt{Feedback Loop}, applied to \textsc{o4-mini} and \textsc{Llama3.3}. Across all strategies, fixed bugs consistently exhibit lower divergence than unfixed bugs. The mean divergence for fixed bugs ranges from $0.28$ to $0.40$ (median: $0.28$–$0.39$), while unfixed bugs display substantially higher scores (mean: $0.49$–$0.56$; median: $0.45$–$0.53$), indicating that greater heterogeneity within hunks correlates with repair failure. Among the evaluated strategies, \texttt{Feedback Loop} achieves the lowest divergence values for fixed bugs in both models (mean: $0.29$–$0.30$; median: $0.30$–$0.31$), and the highest percentage of repairs in the \textsc{Nucleus} and \textsc{Cluster} classes. This pattern suggests improved effectiveness on bugs involving spatially localized edits. 
% In contrast, \texttt{File Context} yields higher divergence and lower proximity-based accuracy, particularly in proximal classes, implying that overly broad contextual input may introduce irrelevant or misleading information. 
These results demonstrate that prompting strategy significantly influences both semantic coherence and spatial locality of successful patches. Notably, the \texttt{Feedback Loop} configuration demonstrates the highest effectiveness in achieving low-divergence repairs within high-proximity contexts, thereby facilitating more precise and coordinated multi-hunk program repair. 

% \ali{we need to talk about the fix percentages in the two tables for spatial proximity classes. We should highlight what the results are.}
The association between spatial proximity and repair success is first evident in Table~\ref{tab:rq3-fixed-not-fixed}. Repair success is highest in the \textsc{Nucleus} and \textsc{Cluster} classes, where hunks are in closer proximity, while substantially lower fix rates are observed in the \textsc{Sprawl} and \textsc{Fragment} categories, with the latter exhibiting no successful repairs. This trend persists across prompt variants, as shown in Table~\ref{tab:rq3-solved-not-solved-prompt-variants}. Among the evaluated strategies, \texttt{Feedback Loop} consistently achieves superior accuracy in close proximity. For example, \textsc{o4-mini} with \texttt{Feedback Loop} repairs 64.41\% of bugs in the \textsc{Nucleus} class and 33.51\% in \textsc{Cluster}, compared to only 22.00\% and 27.27\% in \textsc{Sprawl} and \textsc{Fragment}, respectively. A comparable pattern is observed for \textsc{Llama3.3}.

\begin{figure}[h]
    \centering
    \includegraphics[width=\columnwidth]{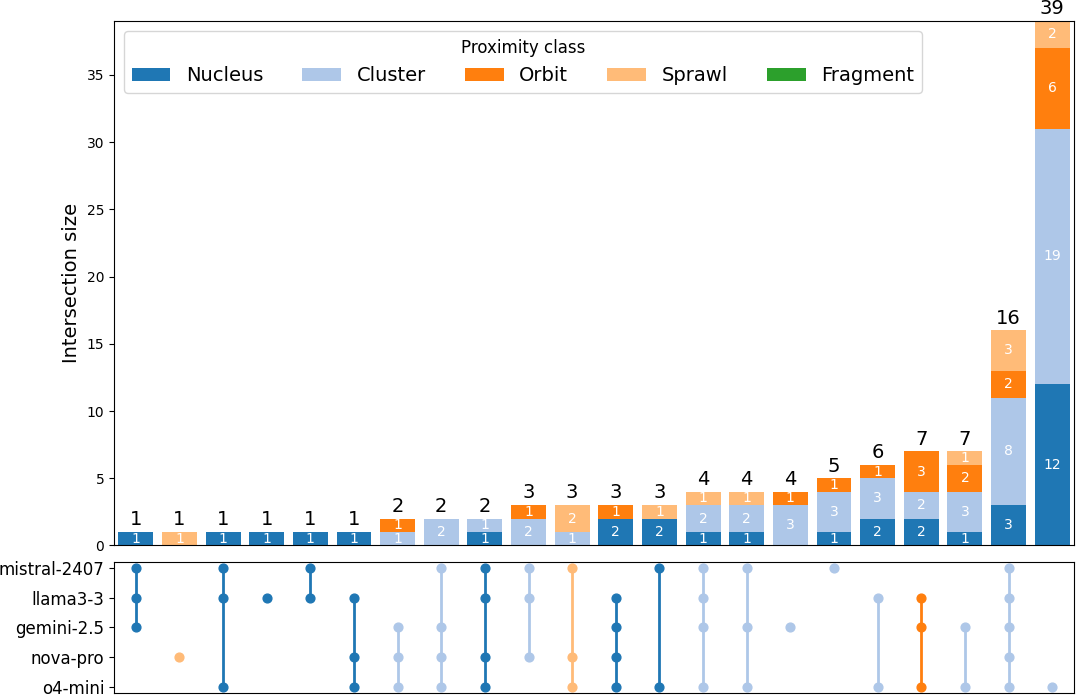}
    % \caption{Upset-style plot of bugs correctly repaired by the best model from each LLM family, grouped by spatial proximity class. Bars indicate model intersections; color segments represent proximity classes.}
    \caption{Upset-style plot showing bugs correctly repaired by the selected models, grouped by spatial proximity class. Bars indicate intersections among models; color segments denote proximity categories.}
    \label{fig:proximity-upset}
\end{figure}

% \ali{this sounds like all 372 bugs were resolved!}
To further investigate whether the spatial proximity of hunks influences the success of LLM-based repair, we analyze the subset of bugs that were correctly fixed by at least one top-performing model from five distinct LLM families. Out of 372 multi-hunk bugs in \benchmarkdataset, a total of 116 were successfully repaired by at least one model. Figure~\ref{fig:proximity-upset} presents an UpSet-style visualization of these 116 fixed bugs, grouped by spatial proximity class and by the specific combinations of models that produced successful patches. Each bar along the horizontal axis represents a unique combination of models (denoted by filled circles in the matrix below), and bars are ordered from left (representing the fewest bugs) to right (representing the most bugs). Within each bar, segments are color-coded by proximity class, \textsc{Nucleus} (dark blue), \textsc{Cluster} (light blue), \textsc{Orbit} (orange), \textsc{Sprawl} (peach), and \textsc{Fragment} (green) with each segment labeled by the exact bug count it represents. The total number of bugs in each model intersection is annotated above the corresponding bar. The lower panel presents a matrix of filled circles and vertical connectors: each row denotes one of the 5 models (from top to bottom: \textsc{Mistral-2407}, \textsc{LLaMA3.3}, \textsc{gemini-2.5-flash}, \textsc{Nova-Pro}, \textsc{o4-mini}), and each column corresponds to the same model combination that generated the bar above. A filled circle indicates that the model participated in that combination; connectors link the participating models.

% \keheliya{Should we say "just the llm alone could not fix Fragment category", because if augmented, some of these can be fixed based on table vii.}
Two trends are evident. First, the majority of successful repairs fall into the \textsc{Nucleus} and \textsc{Cluster} categories, which represent cases where the relevant code elements are tightly co-located. These proximity classes dominate across both shared and model-specific intersections. Second, cases in the \textsc{Orbit} and \textsc{Sprawl} categories, where relevant context is more dispersed, are less frequently solved. 
The LLM alone was unable to fix any bug in the \textsc{Fragment} category, which reflects extreme dispersion or disconnection of relevant context (Table~\ref{tab:rq3-fixed-not-fixed}).
The largest set of bugs (39 bugs) is uniquely attributable to \textsc{o4-mini}, while the second-largest (16 bugs) was solved by all five models. Smaller intersections reveal model-specific capabilities or divergences, particularly in handling less proximal context.

Collectively, these findings suggest that successful LLM-based repairs for multi-hunk bugs are characterized by (1) semantically coherent edits with low hunk divergence and (2) spatially localized dependencies with high proximity. %These trends support the validity of both metrics as indicators of multi-hunk repair difficulty. % and motivate future approaches that incorporate divergence-aware prompt construction.

% - Run: `wilcox.test(avg_divergence ~ solved, data = df_best_o4mini)`
% - Visualize using boxplot with p-value annotated in title.
% - If p < 0.05, conclude lower divergence significantly correlates with successful repair.
% - Integrate into Figure X and describe in Section III.F (RQ3 results).

% \nashid{TODO: (a) descriptive statistics of 16 bugs that best performing LLM from each family can solve, (b) descriptive statistics for the rest of the 356 bugs, common solved, and then best performing model from each family. And finally, not solvable by any model.}
% 
% I have not done it, as there is no space to add it
% 
% 
% \subsection{Divergence, Proximity, and Repair Difficulty (RQ4)}
\subsection{Divergence, Proximity, and Repair Difficulty (RQ4)}
% We now examine whether hunk divergence and spatial proximity can serve as indicators of LLM repair difficulty. Here, we treat repair difficulty as the extent to which LLMs fail to produce a correct fix. This analysis focuses on two core factors: hunk divergence and spatial proximity.
We now investigate whether hunk divergence and spatial proximity function as predictive indicators of LLM repair difficulty, i.e., the likelihood that a model fails to synthesize a correct patch. This analysis considers two dimensions of multi-hunk patches: the degree of intra-patch variability (hunk divergence) and the extent of spatial dispersion across the program hierarchy (spatial proximity).

% To assess this, we conduct a Wilcoxon rank-sum test comparing the divergence scores of fixed and not fixed bugs for the top-performing model from each LLM family. Across all five models, fixed bugs exhibit significantly lower divergence (mean: $0.26$–$0.28$; median: $0.24$–$0.29$), while not fixed bugs show higher divergence (mean: $0.49$–$0.56$; median: $0.45$–$0.53$). All tests yield statistically significant differences ($p < 10^{-5}$), confirming that high divergence is strongly associated with repair failure.

% moved text from RQ3 to RQ4 - START
% \keheliya{isn't it "two-sample"?}
To assess this, we first partitioned the data for the Top-5 LLMs into two independent samples, fixed and unfixed bugs, and applied a two-sample Wilcoxon rank-sum test to their divergence scores. Across all five top-performing models, the divergence distributions for fixed and unfixed bugs differ significantly. All Wilcoxon tests yield $p$-values below $10^{-4}$, with three models reporting $p < 10^{-7}$ and the largest $p$-value at $2.6\times10^{-5}$. These results confirm that higher divergence is strongly associated with failure to generate a correct patch. 
% We also conducted Cliff's Delta test to observe the effect sizes. \texttt{o4-mini} reported $0.533$ for basic prompting, and $0.554$ for feedback-loop prompting. These numbers are greater than the threshold of $0.474$, indicating large differences between the solved and unsolved groups by hunk divergence.
We also computed Cliff’s Delta to assess the effect sizes.
Under LLM-only and feedback-loop prompting, the delta values for \textsc{o4-mini} were $0.533$ and $0.554$, respectively.
Both values exceed the threshold of $0.474$, indicating large effect sizes, that is, substantial differences in hunk divergence between successfully and unsuccessfully repaired bugs.
% \keheliya{Can we report on effect sizes too using Cliff’s delta? Interpret them as follows: negligible when \textit{delta} $<$ 0.147 , small when 0.147 $\leq$ \textit{delta} $<$ 0.33 , medium when 0.33 $\leq$ \textit{delta} $<$ 0.474 , and large otherwise.}
% The consistent statistical significance across models indicates that divergence is a meaningful signal of patch complexity and LLM repair performance. 
The consistent statistical significance observed across models suggests that hunk divergence may serve as a meaningful indicator of patch complexity and the effectiveness of LLM-based repair.

% Successful repairs are consistently associated with low divergence across all five models (mean $\approx 0.26$–$0.28$; median  $\approx 0.24$–$0.29$). Variance remains narrow (standard deviation $< 0.20$), indicating high semantic alignment across generated patches. The Wilcoxon tests ($p < 10^{-5}$) further confirm that divergence scores for solved bugs are significantly lower than those of unsolved ones. For the five LLM families, every Wilcoxon rank-sum test rejects the null hypothesis: the largest $p$-value is $2.6\times10^{-5}$, and three of the five values fall below $10^{-7}$. 

% p-value discussion for prompt variants as well
We observe similar trends when comparing augmented prompting configurations (Table~\ref{tab:rq3-solved-not-solved-prompt-variants}). Among the six prompting variants, four yield $p$-values below $10^{-7}$, one yields $p = 3.8\times10^{-3}$, and all remain below the conventional threshold of $0.05$. This demonstrates that the distinction between fixed and unfixed bugs based on divergence holds across different prompt strategies, further validating its predictive value.

% For the six augmented prompting strategies, the same pattern holds: four variants yield $p<10^{-7}$, one variant gives $p=3.8\times10^{-3}$, while the weakest still remains below $2\times10^{-2}$. These results show that, whether we vary the underlying model or the augmented prompting technique, the divergence distributions of solved and unsolved bugs are statistically distinct, with the strongest effects concentrated in the lowest-divergence region.

% \ali{I like the Upset-style plot for spatial proximity analysis. However, I think we are not doing a proper analysis for hunk divergence. Since hunk divergence is a continuous, quantitative measure (unlike the categorical proximity classes), there are several effective ways to visualize its value and relationship to LLM repair success: Histogram or KDE Plot (Distribution of Divergence Scores), Scatter plots, heatmaps, buckets of divergence versus success/failure, Bin divergence scores (e.g., 5–7 buckets), etc}
% 
% \ali{do we have the individual patch divergence scores and outcomes (successfully repaired or not)? I can also play with the visualization if you like.}
%
% Thanks a lot for your help with visualization.
% Link to readme with the details for patch divergence scores and outcomes:https://github.com/testcue/birch/blob/main/redwood/hunk_divergence_v4_bleu/README.md
% I have shared the readme in slack, so closing the comment here for the time-being.

Figure~\ref{fig:llm_divergence_violin} presents faceted violin plots showing the distribution of average hunk divergence scores for each model, separated by outcome. Each violin illustrates the density of divergence values, while the embedded boxplot indicates the median and interquartile range, capturing the central 50\% of the data. Fixed bugs are consistently associated with lower divergence scores across all models. This clear and consistent pattern demonstrates that hunk divergence is a discriminative signal for reasoning about multi-hunk bug complexity and the success of LLM-based repair.
% moved text from RQ3 to RQ4 - END

\begin{figure}%[h]
    \centering
    \includegraphics[width=\columnwidth]{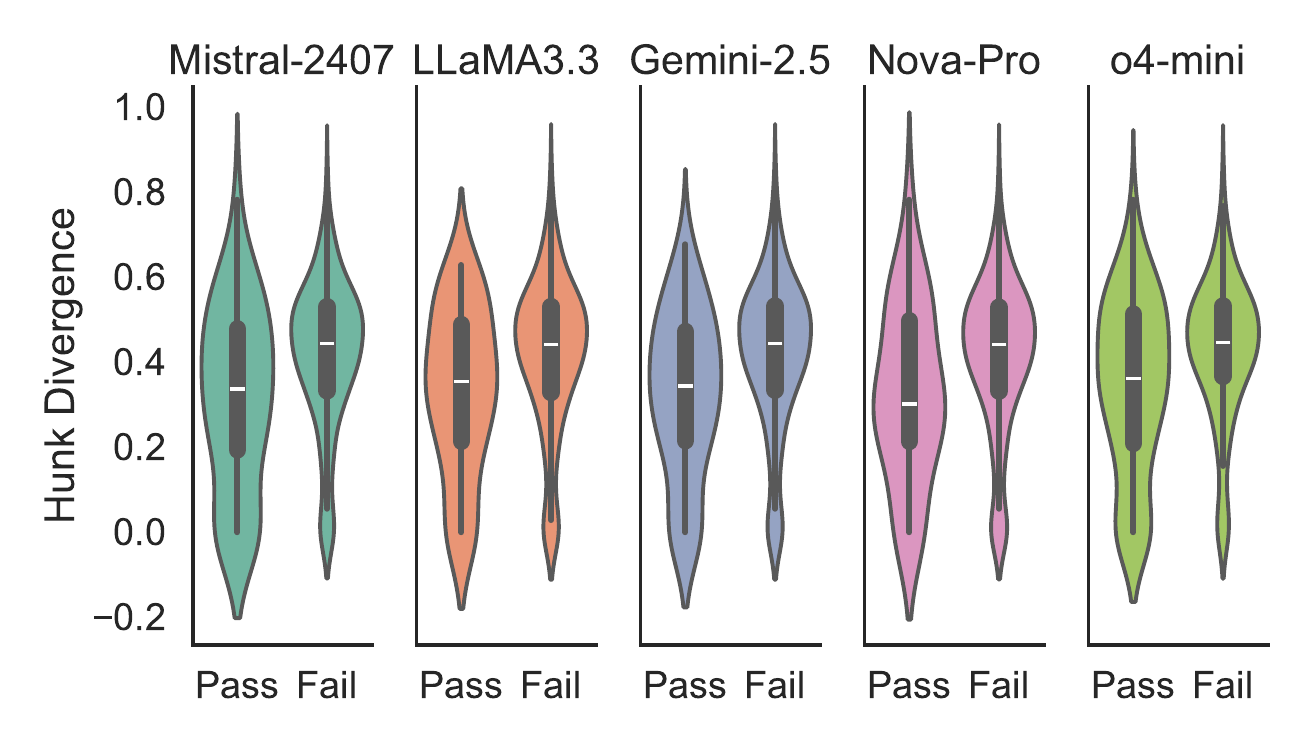}
    \caption{Faceted violin plots of average hunk divergence, grouped by model and split by outcome.}
    \label{fig:llm_divergence_violin}
\end{figure}

In addition to divergence, spatial proximity also emerges as a strong categorical predictor. As shown in Figure~\ref{fig:proximity-upset},  repair success declines consistently from \textsc{Nucleus} to \textsc{Sprawl}, with the \textsc{Fragment} class remaining entirely unsolved across all models. This trend suggests that as fault-relevant context becomes more dispersed, LLMs are increasingly unable to synthesize correct patches. The complete absence of fixes in the \textsc{Fragment} category indicates a hard boundary for current model capabilities in handling non-local, scattered dependencies.

Our analysis shows that LLMs exhibit a significant drop in repair accuracy as hunk divergence increases and spatial proximity decreases. Bugs with high intra-patch variation or dispersed hunks across methods, files, or packages remain largely unrepaired. These findings underscore the need for models and prompts that explicitly account for lexical and structural dispersion in multi-hunk patches. 

\section{Threats to Validity}

% While \hunkdivergence captures lexical, structural, and file-level separation, it may not fully reflect deeper semantic dependencies such as control flow or variable usage across methods or files.

% \header{Dataset}

% previous
% \benchmarkdataset is derived exclusively from Java projects, which may limit generalizability to other languages. Although these bugs are real-world and diverse, their distribution reflects the design and testing philosophy of Java projects, which may bias the scope of repair challenges.

% after addressing the meta-review comment
% \change
\benchmarkdataset is derived exclusively from Java projects, which may limit generalizability to other languages. Although these bugs are real-world and diverse, their distribution reflects the design and testing philosophy of Java projects, which may bias the scope of repair challenges. We evaluate on Java because it is a popular language and use the widely adopted Defects4J benchmark. Our metrics, hunk divergence and spatial proximity, and our evaluation steps do not rely on language-specific features, so the underlying insights should apply to other languages; nevertheless, additional experiments on non-Java datasets are required to empirically validate this hypothesis and assess generalizability beyond Java and Defects4J.

Our goal is to study how the structure of real developer-written patches affects LLM repair. We use the original patches from Defects4J because they reflect how developers fixed the bugs. A simpler or semantically equivalent patch may exist for some bugs; however, we intentionally compute hunk divergence and spatial proximity from the actual developer patch, not hypothetical alternatives. This provides us with a consistent and realistic approach to studying what makes multi-hunk repair challenging.
% \stopchange{}

% \header{Evaluation Metrics} 
We use the Plausible@1 metric to reduce inference cost, which may understate model accuracy in iterative settings. Decoding parameters and prompt templates are fixed across models to ensure fairness, potentially favoring some architectures. However, our aim is comparative analysis under uniform conditions and not model-specific optimization.

% \header{Fault Localization Assumption} 
% \keheliya{is this the correct table here? based on that table we can fix many fragment bugs}
Our evaluation assumes perfect fault localization, following prior LLM-based repair studies to isolate patch generation quality~\cite{xia:alpharepair:fse22, fan:code-repair-with-llm:icse23, iter, repair-agent, cedar, glance, reptory, katana}. While this sets an upper bound on accuracy, real-world scenarios often involve imperfect or incomplete localization. Notably, even with perfect localization, LLMs struggle with multi-hunk repairs, failing on all Fragment-class bugs (Table~\ref{tab:rq3-fixed-not-fixed}), highlighting challenges in semantic coordination beyond localization.
% We assume perfect fault localization, following prior repair studies~\cite{xia:alpharepair:fse22, fan:code-repair-with-llm:icse23, iter, repair-agent, cedar, glance, reptory, katana}, to isolate patch generation quality. While this sets an upper bound on accuracy, LLMs still fail on all Fragment-class bugs (Table~\ref{tab:rq3-fixed-not-fixed}), underscoring challenges in coordinating scattered edits even with ideal localization.

% \header{Data leakage} 
% \ali{repeating myself here: performance is about runtime in SE, not accuracy! Don't use performance to mean accuracy (ever) in SE papers.}
% \ali{is data leakage really important in this work?}
% Since our dataset consists of publicly available open-source projects, we cannot rule out the possibility that parts of the dataset were included in the pretraining data of the LLMs. Such data leakage could influence accuracy. However, our objective is not to evaluate memorization or out-of-distribution generalization, but to assess the effectiveness of LLMs on realistic, complex multi-hunk bugs. In practical development settings, reliance on prior knowledge is both inevitable and appropriate. Moreover, the relatively low repair accuracy we observed indicates that the models do not merely reproduce memorized bug-fix pairs. 

% \header{Model and API Variance} 
LLMs are accessed via third-party APIs, and execution can vary due to backend updates or system-level nondeterminism. Although decoding is configured to be deterministic (temperature = 0), minor variations may still occur. We mitigate this by logging all inputs, outputs, and execution metadata for reproducibility.

% \header{Toolchain Bias} 
Our implementation uses \texttt{JavaParser} to extract ASTs and compute divergence metrics such as node-level distance and tree diameter. Alternative parsers (such as \texttt{javalang}) may yield slight variations due to differences in AST tree construction, though all models are evaluated under the consistent criteria.

% \header{Reproducibility} 
To support reproducibility, we release \benchmark, all scripts, prompts, and the \benchmarkdataset dataset~\cite{anon-repo}.
\section{Related Work}

Most APR benchmarks offer limited support for multi-hunk bugs. While Defects4J~\cite{defects4j}, Bugs.jar~\cite{bugsjar:msr18}, BugsInPy~\cite{bugsinpy:fse20}, and LMDefects~\cite{fan:code-repair-with-llm:icse23} include them, others offer limited coverage~\cite{bears:saner19, manybugs:tse15, bugsjs} or focus on single-hunk edits~\cite{chen:codrep:arxiv18, lin:quixbugs:splash17, karampatsis:manysstubs4j:msr20}. Existing multi-hunk LLM repair studies rely on synthetic~\cite{fan:code-repair-with-llm:icse23}, function-level~\cite{srepair:arxiv24}, or algorithmic~\cite{yang:fine-tuning-for-repair-with-llm:arxiv24} benchmarks, limiting generalizability. We introduce \textsc{Hunk4J}, a real-world, multi-hunk benchmark derived from Defects4J, augmented with divergence metrics for evaluating LLM-based repair.

% Several studies have examined software change complexity using hunk-level and structural metrics. Graves et al.\cite{graves2000predicting} introduced change burst metrics, including hunk and file counts, for fault prediction. D’Ambros et al.\cite{dambros2012evaluating} used change coupling and scatter as predictors of defect-prone changes. Kim et al.\cite{kim2008classifying} showed that fixes spanning multiple files are more error-prone. Our work differs by formalizing hunk divergence and spatial proximity as novel metrics to quantify intra-patch complexity and empirically analyze their impact on LLM-based program repair.

% Prior studies have used change metrics to analyze fault-proneness. Graves et al.\cite{graves2000predicting} focused on coarse-grained indicators such as file and modification counts. D’Ambros et al.\cite{dambros2012evaluating} demonstrated the predictive value of change coupling and scatter. Ferzund et al.\cite{ferzund2009software} introduced hunk-level metrics for bug prediction. Unlike these works, which target fault prediction, we focus on quantifying semantic and structural heterogeneity within multi-hunk patches to assess the effectiveness of LLM-based repair.
Prior studies have applied change metrics for fault prediction. Graves et al.~\cite{graves2000predicting} used coarse-grained measures such as file and modification counts; D’Ambros et al.~\cite{dambros2012evaluating} highlighted the predictive power of change coupling and scatter; and Ferzund et al.~\cite{ferzund2009software} proposed hunk-level metrics. In contrast, we quantify semantic and structural heterogeneity in multi-hunk patches to assess LLM-based repair effectiveness.

Prior work on multi-hunk repair with LLMs~\cite{fan:code-repair-with-llm:icse23, srepair:arxiv24, yang:fine-tuning-for-repair-with-llm:arxiv24, repair-agent, huang:empirical-fine-tuning-for-apr:ase23, xin:empirical-study-hunk:fse24, xia:alpharepair:fse22} exhibits wide variation in datasets, prompting strategies, and evaluation protocols, often reporting aggregate results without isolating multi-hunk accuracy. While prior studies have examined patch properties such as hunk divisibility~\cite{xin:empirical-study-hunk:fse24}, they do not account for spatial dispersion or intra-patch heterogeneity. We address this limitation by introducing \emph{Hunk Divergence} and \emph{Spatial Proximity}, two metrics that capture lexical, structural, and hierarchical relationships between hunks to enable complexity-aware analysis.

\section{Conclusion}

We introduce the first characterization of multi-hunk bugs through the lens of intra-patch divergence. 
Our metric, \hunkdivergence, captures lexical, structural, and file-separation dissimilarity between hunks, while \proximityclass classifies their distribution across the codebase. 
% To enable systematic evaluation, we introduce \benchmark, the first benchmark specifically designed for multi-hunk repair.
% Our findings reveal that current models are ill-equipped to reason about disconnected yet semantically related code regions.
Our findings reveal that multi-hunk bugs exhibit substantial variability in divergence, with no single model achieving consistently high accuracy across this spectrum.
Collectively, these contributions lay the foundation for developing \emph{divergence-aware} techniques capable of addressing the unique challenges posed by multi-hunk bugs. 
In future work, we aim to develop a divergence-aware model selector that routes bugs to the most capable and cost-efficient LLM based on predicted repair complexity.
% In particular, we This selective model selection may improve both cost, computational efficiency and repair effectiveness by aligning model choice with the predicted divergence of the bug.
% Future work includes developing divergence-aware model selection strategies that route bugs to appropriate LLMs based on estimated fix dispersion, and extending divergence metrics to tasks such as Github issue resolution.
% Looking forward, repair systems needs to incorporate divergence-aware prompt design and context retrieval mechanisms that can localize, retrieve, and jointly modify semantically coupled hunks spread across files in the codebase.

\balance
\bibliographystyle{IEEEtran}
\interlinepenalty=10000
\bibliography{references}

\end{document}